\documentstyle[11pt]{article}
\title{Some Properties of Noether Charge and a Proposal
for Dynamical Black Hole Entropy}
\author{Vivek Iyer and Robert M. Wald\\Enrico Fermi Institute and
Department of Physics\\
University of Chicago\\5640 S. Ellis Ave.\\Chicago, IL 60637}
\date{March 15 1994}

\begin{document}

\maketitle

\newcommand{\be}{\begin{equation}}
\newcommand{\ee}{\end{equation}}
\newcommand{\bea}{\begin{eqnarray}}
\newcommand{\eea}{\end{eqnarray} }
\newcommand{\beas}{\begin{eqnarray*}}
\newcommand{\eeas}{\end{eqnarray*} }
\newcommand{\bdm}{\begin{displaymath}}
\newcommand{\edm}{\end{displaymath} }

\newcommand{\dee}{\partial}
\newcommand{\eps}{\mbox{{\boldmath $\epsilon$}}}
\newcommand{\Om}{\Omega}
\newcommand{\om}{\mbox{{\boldmath $\omega$}}}
\newcommand{\k}{\xi}
\newcommand{\kt}{\tilde{\xi}}
\newcommand{\de}{\delta}
\newcommand{\kap}{\kappa}
\newcommand{\cd}{\cdot}
\newcommand{\Liek}{{\cal L}_\k}
\newcommand{\Liekt}{{\cal L}_{\kt}}
\newcommand{\Liet}{{\cal L}_t}
\newcommand{\th}{{\bf \Theta}}
\newcommand{\bB}{{\bf B}}
\newcommand{\bS}{{\bf S}}
\newcommand{\bT}{{\bf T}}
\newcommand{\bU}{{\bf U}}
\newcommand{\bL}{{\bf L}}
\newcommand{\bJ}{{\bf J}}
\newcommand{\bK}{{\bf K}}
\newcommand{\bQ}{{\bf Q}}
\newcommand{\bE}{{\bf E}}
\newcommand{\bW}{{\bf W}}
\newcommand{\bX}{{\bf X}}
\newcommand{\bXt}{{\bf \tilde{X}}}
\newcommand{\bY}{{\bf Y}}
\newcommand{\bZ}{{\bf Z}}
\newcommand{\bff}{{\bf f}}
\newcommand{\bmu}{\mbox{\boldmath $\mu$}}

\newcommand{\gammao}{\stackrel{\circ}{\gamma}}
\newcommand{\gammaoo}{\stackrel{\circ}{\gamma}_o}
\newcommand{\odel}{\del_{a_1}}
\newcommand{\osdelo}{\delo_{(a_1}}


\newcommand{\del}{{\bf \nabla}}
\newcommand{\delo}{\stackrel{\circ}{\bf \nabla}}

\newcommand{\odelo}{\delo_{a_1}}
\newcommand{\pdelo}{\delo_{(a_1}...\delo_{a_p)}}
\newcommand{\kdelo}{\delo_{(a_1}...\delo_{a_k)}}
\newcommand{\ldelo}{\delo_{(a_1}...\delo_{a_l)}}
\newcommand{\sdelo}{\delo_{(a_1}...\delo_{a_s)}}
\newcommand{\smodelo}{\delo_{(a_1}...\delo_{a_{s-1})}}
\newcommand{\smtdelo}{\delo_{(a_1}...\delo_{a_{s-2})}}

\newcommand{\ldel}{\del_{(a_1}...\del_{a_l)}}
\newcommand{\mdel}{\del_{(a_1}...\del_{a_m)}}
\newcommand{\idel}{\del_{(a_1}...\del_{a_i)}}
\newcommand{\smtdel}{\del_{(a_1}...\del_{a_{s-2})}}
\newcommand{\tidel}{\del_{(a_2}...\del_{a_i)}}

\newcommand{\lsodelo}{\delo_{(a_1}}
\newcommand{\lspdelo}{\delo_{(a_1}...\delo_{a_p}}
\newcommand{\lssmodelo}{\delo_{(a_1}...\delo_{a_{s-1}}}
\newcommand{\lsidelo}{\delo_{(a_1}...\delo_{a_i}}

\newcommand{\nsidel}{\del_{a_1}...\del_{a_i}}
\newcommand{\nstidel}{\del_{a_2}...\del_{a_i}}


\newtheorem{theorem}{Theorem}[section]
\newtheorem{lemma}{Lemma}[section]
\newtheorem{proposition}{Proposition}[section]
\newtheorem{defn}{Definition}[section]

\begin{abstract}
We consider a general, classical theory of gravity with arbitrary
matter fields in $n$ dimensions,
arising from a diffeomorphism invariant Lagrangian, $\bL$.
We first show that $\bL$ always can be
written in a ``manifestly covariant" form. We then show that
the symplectic potential current $(n-1)$-form, $\th$, and the
symplectic current $(n-1)$-form, $\om$, for the theory
always can be globally defined in a covariant manner.
Associated with any infinitesimal diffeomorphism is a Noether current
$(n-1)$-form, $\bJ$, and corresponding Noether charge
$(n-2)$-form, $\bQ$. We derive a general ``decomposition formula" for
$\bQ$. Using this formula for the Noether charge, we prove
that the first law of black hole mechanics holds for arbitrary perturbations
of a stationary black hole. (For higher derivative theories, previous
arguments had established this law only for stationary
perturbations.)  Finally, we propose a local, geometrical prescription
for the entropy, $S_{dyn}$, of a dynamical black hole. This prescription agrees
with the Noether charge formula for stationary black holes and their
perturbations, and is independent of all ambiguities associated with the
choices of $\bL$, $\th$, and $\bQ$. However, the issue of whether
this dynamical entropy in general obeys a ``second law" of black hole
mechanics remains open. In an appendix, we apply
some of our results to theories with a nondynamical metric and also
briefly develop the theory of stress-energy pseudotensors.

{\bf PACS \#: } 04.20.-q, 0.4.20.Fy, 97.60.Lf
\end{abstract}
\newpage

\oddsidemargin=0in
\evensidemargin=0in
\topmargin=-.5in
\textwidth=6.7in
\textheight=9in
\hsize=6.27truein

\begin{section}{Introduction}

Recently, many authors have investigated the validity of the
first law of
black hole mechanics and the definition of the entropy of a black hole
in a wide class of theories derivable from a
Hamiltonian or Lagrangian \cite{SW}-\cite{CT}.
In particular, in \cite{W1} the first law was proven to hold
in an arbitrary theory of gravity derived from a diffeomorphism
invariant Lagrangian, and the quantity playing the role of the entropy
of the black hole
was identified as the integral over the horizon of the
Noether charge associated with the horizon Killing vector field. Although
some key issues concerning the validity of the first law and the definition
of black hole entropy in a general theory of gravity were thereby
resolved, the analysis of \cite{W1}, nevertheless, was
deficient in the following ways: (1) It
was not recognized that a diffeomorphism covariant choice of the
symplectic potential current form always can be made.
Consequently, several
steps in the arguments were made in an unnecessarily awkward manner.
(2) While a completely general proof of the first law of black hole
mechanics was given for perturbations to nearby stationary black holes,
a proof of the first law for non-stationary perturbations was given only
for theories in which the Noether charge takes a particular, simple
form. (3) A proposal was made for defining the entropy of a dynamical
black hole. However, this proposal made use of a rather arbitrary choice
of algorithm for defining the
symplectic potential current form, and it turns out to
possess the undesireable feature that the addition of an
exact form to the Lagrangian (which has no effect upon the equations of
motion of the theory) can induce a nontrivial change in this proposed
formula for the entropy of a dynamical black hole \cite{JM2}.

The main purposes of this paper are to remedy
all of the above deficiencies, and, in addition, develop further
the theory of Noether currents and charges in diffeomorphism invariant
theories. We shall
show, first, that the Lagrangian of a diffeomorphism invariant theory
always can be expressed in a manifestly covariant form. This will enable
us to give globally defined, covariant definitions of the symplectic potential
current form, $\th$, and symplectic current form, $\om$, in an arbitrary
diffeomorphism invariant theory. Furthermore, results on the general
form of $\th$ in an arbitrary theory will be obtained, from which it will
follow that the Noether charge form, $\bQ$, always has a particular,
simple structure. As a consequence of this structure of $\bQ$,
the first law of black hole mechanics will be proven to hold
for nonstationary perturbations in an arbitrary theory of gravity derived
from a diffeomorphism covariant Lagrangian. We then shall propose
a definition of the entropy, $S_{dyn}$, of an arbitrary
cross-section of a nonstationary black hole, wherein $S_{dyn}$ is given
by an integral over the horizon of a local, geometrical
quantity. Our proposed
definition agrees with the known answer (as determined by the
first law) for stationary black holes and their
perturbations, and is independent of all ambiguities associated with the
choices of $\bL$, $\th$, and $\bQ$. However, it is not known whether our
$S_{dyn}$ obeys a ``second law" in general theories of gravity.
The paper concludes with an Appendix
in which some of our results are applied to theories with a nondynamical
metric, and some results on stress-energy pseudotensors are obtained.

We shall follow the notation and conventions of \cite{Wald}. All
spacetimes, tensor fields, and surfaces considered in this paper will
be assumed to be smooth ($C^\infty$).

\end{section}

\begin{section}{The form of the Lagrangian for diffeomorphism invariant
theories}
We wish to consider, here, Lagrangian theories on
an $n$-dimensional, oriented
manifold $M$, with the dynamical fields consisting
of a Lorentz signature metric $g_{ab}$, and other
fields $\psi$. For simplicity and definiteness, we shall restrict
consideration to the case where
$\psi$ is a collection of tensor fields on $M$
(with arbitrary index structure). However, we foresee no essential
difficulty in extending our analysis and results to the case where
$\psi$ is a section of an arbitrary vector bundle which
possesses a connection uniquely determined by $g_{ab}$.

We start with the general form of a
Lagrangian postulated in \cite{LW} and \cite{W1}: Specifically, we
introduce an arbitrary, fixed, globally defined, derivative operator, $\delo$,
and take the Lagrangian to be a function of the
quantities $g_{ab}$, $\psi$, and finitely many of their symmetrised
derivatives with respect to $\delo$.
In addition, the Lagrangian is permitted to depend on additional
``background fields" $\gammao$ -- which, like $\delo$, do not change
under variation of the dynamical fields; a good example of such a
background field upon which the Lagrangian could depend is
the curvature, ${\stackrel{\circ}{R}_{bcd}}{}^e$, of $\delo$.  Thus,
we take the Lagrangian to be an $n$-form locally
constructed -- in the precise sense explained in \cite{W2} --
out of the following quantities,
\be
\bL =  \bL \left(g_{ab},\odelo g_{ab}, ...,\kdelo g_{ab},
\psi,  \odelo \psi, ...,\ldelo \psi,\gammao\right),
\label{lagr1}
\ee
Here and in what follows, we use boldface letters to denote
differential forms on spacetime, and we shall, in general, suppress their
tensor indices.
In the following, we also shall collectively refer to the
dynamical fields ``$\psi$ and $g$"
as ``$\phi$".

We shall be concerned here only with diffeomorphism invariant
theories, i.e., the Lagrangian will be assumed to be diffeomorphism
covariant in the sense that
\be
\bL (f^*(\phi)) = f^* \bL (\phi),
\label{lagr2}
\ee
where $f^*$ is the action induced on the fields by a
diffeomorphism $f: M \rightarrow M$. Note that on the left side of this
equation
$f^*$ does not act on $\delo$ or the background fields $\gammao$.

The main result to be established
in this section is that any Lagrangian, $\bL$,
which is diffeomorphism covariant in the sense of eq.(\ref{lagr2})
always can be written in a manifestly covariant form. More precisely,
we have the following lemma, which is closely related to
``Thomas replacement theorem" \cite{Th}:
\begin{lemma}
If $\bL$ as given in (\ref{lagr1}) is diffeomorphism covariant
in the sense of (\ref{lagr2}) then L can be re-expressed as
\be
\bL = \bL \left(g_{ab}, \del_{a_1}R_{bcde},...,
\del_{(a_1}...\del_{a_m)}R_{bcde},
\psi, \del_{a_1}\psi, \del_{(a_1}...\del_{a_l)} \psi\right)
\label{newl}
\ee
where $\del$ denotes the derivative operator associated with $g_{ab}$,
 $m = \max (k-2, l-2)$,
$R_{abcd}$ denotes the curvature of $g_{ab}$, and the absence of any
dependence on ``background fields" in (\ref{newl}) should be noted.
\end{lemma}
{\em Proof.} We begin by using the relation (written here schematically)
$$\delo \alpha = \del \alpha+\alpha \cd \mbox{terms linear in}
\delo g$$ for any tensor field, $\alpha$,
to re-write all of the $\delo$-derivatives of the matter
fields, $\psi$, in terms of $\del$-derivatives of $\psi$ -- where $\del$
is the derivative operator associated with $g$ -- together
with terms involving
the $\delo$-derivatives of $g$. Next, we re-write the $\del$-derivatives
of $\psi$ in terms of symmetrized $\del$-derivatives and the curvature of
$g$ and its derivatives. Then we re-write the curvature of $g$ and
its derivatives in terms of
$\delo$-derivatives of $g$ and the curvature of
$\delo$ and its $\delo$-derivatives. Finally, we write all of the
$\delo$-derivatives of $g$ in terms of symmetrized $\delo$-derivatives
of $g$ and the curvature of
$\delo$ and its $\delo$-derivatives. We thereby obtain
\be
\bL = \bL \left(g, \odelo g_{ab}, ..., \sdelo g_{ab},
\psi, \odel \psi, ...,\ldel \psi, \gammao{}'\right)
\ee
where $s = \max(k,l)$ and
$\gammao{}'$ is comprised by $\gammao$ together with
the curvature of $\delo$ and (finitely
many of) its $\delo$-derivatives.
Next we eliminate $\delo_a g_{bc}$ and its higher $\delo$-derivatives
in favor of
\be
{C^e}_{cd}=\frac{1}{2}g^{ef}(\delo_c g_{fd} + \delo_d g_{fc} - \delo_f g_{cd})
\ee
and its $\delo$- derivatives via the substitution
\be
\delo_a g_{bc} = g_{ec} {C^e}_{ab} + g_{be} {C^e}_{ac}
\ee
Again, we express all $\delo$-derivatives of ${C^e}_{cd}$ in terms of
symmetrized $\delo$-derivatives and the curvature of $\delo$.
We thereby obtain
\be
\bL = \bL \left(g, {C^e}_{cd}, \odelo {C^e}_{cd},...,\smodelo {C^e}_{cd},
\psi, \odel \psi, \ldel \psi, \gammao{}'\right).
\ee
It is tedious but straightforward to check that the symmetrized
derivatives of $C$ can be re-written as
\bea
\pdelo {C^e}_{cd}
&=& \lspdelo {C^e}_{cd)} \nonumber \\
&&+ \frac{p+3}{4(p+1)(p+2)} \sum_i
\del_{(a_1}...\hat{\del}_{a_i}...\del_{a_p)} \left({R_{ca_id}}^e
+ {R_{da_ic}}^e \right)\nonumber \\
& & \frac{3p+4}{8p(p+1)(p+2)} \sum_{i \ne j} \left(
\del_{(d}\del_{a_1}...\hat{\del}_{a_i} \hat{\del}_{a_j}...\del_{a_p)}
{R_{a_ica_j}}^e  \right. \nonumber \\
&&\left. +\del_{(c}\del_{a_1}...\hat{\del}_{a_i} \hat{\del}_{a_j}...\del_{a_p)}
{R_{a_ida_j}}^e \right) \nonumber \\
& &+\mbox{terms involving no more than $(p-1)$ $\delo$- derivatives of } C
\label{rearr1}
\eea
where $\hat{\del}_{a_i}$ means the omission of this derivative operator in the
sequence.
By repeatedly making this substitution in sequence, starting with $p = s-1$,
then $p = s - 2$, etc., and, at each step, writing multiple derivatives in
terms
of symmetrized derivatives and curvatures, we can express the Lagrangian as
{\samepage
\bea
\bL&=&\bL\left(g_{ab},{C^e}_{cd},
\osdelo {C^e}_{cd)},..., \lssmodelo {C^e}_{cd)},
 R_{bcde}, \odel R_{bcde},..., \smtdel R_{bcde},\right. \nonumber\\
& & \left. \psi, \odel\psi, ..., \ldel \psi,\gammao{}' \right)
\label{lagr3}
\eea
}
The infinitesimal version of the diffeomorphism covariance condition
(\ref{lagr2}) is
\be
{\cal L}_\k L(\phi) =
\frac{\partial L}{\partial \phi} {\cal L}_\k \phi
\label{diff}
\ee
Applying this to eq.(\ref{lagr3}), we obtain
\bea
&&\frac{\dee \bL}{\dee {C^e}_{cd}} \Liek {C^e}_{cd}
+ \frac{\dee \bL}{\dee \lsodelo {C^e}_{cd)}} \Liek \lsodelo {C^e}_{cd)} +...
\nonumber \\
&&+\frac{\dee \bL}{\dee \lssmodelo {C^e}_{cd)}} \Liek \lssmodelo {C^e}_{cd)}
 + \frac{\dee \bL}{\dee \gammao{}'} \Liek \gammao{}' \nonumber \\
&=&\frac{\dee \bL}{\dee {C^e}_{cd}} \de {C^e}_{cd} +
 \frac{\dee \bL}{\dee \lsodelo {C^e}_{cd)}}
\osdelo \de {C^e}_{cd)} +...\nonumber\\
&&+ \frac{\dee \bL}{\dee \lssmodelo {C^e}_{cd)}} \lssmodelo \de {C^e}_{cd)}
\label{lagr4}
\eea
where
\be
\de {C^a}_{bc} = g^{ad}\left(\delo_{(b} \delo_{c)}\k_{d}
- {{\stackrel{\circ}{R}}_{d(bc)}}{}^e \k_e\right)
-2\delo{}^{(a}\k^{d)}g_{de}C^e_{bc}
\ee
is the variation of ${C^a}_{bc}$ arising from the metric variation
$\de g_{ab} = {\cal L}_\k g_{ab} =2 \del_{(a} \k_{b)}$. (Note that
the terms in eq.(\ref{diff}) arising from the variations of
$\psi$, $g$, and the curvature, $R$, of $g$ cancel and were therefore omitted
when writing eq.(\ref{lagr4})). The dependence of the terms in eq.(\ref{lagr4})
on $\k^a$ and its symmetrized $\delo$-derivatives should be noted:
no more than one derivative of $\k^a$ appears
on the left side of this equation, but the right side contains
terms with as many as $(s+1)$ symmetrized
derivatives of $\k^a$. Since, at any given point in $ M$,
$\k^a$ and its symmetrized derivatives
can be chosen independently,
it follows directly that a necessary condition for eq.(\ref{lagr4}) to
hold for all $\k^a$ is
\be
\frac{\dee \bL}{\dee \lsidelo {C^e}_{cd)}} = 0 \mbox{, for } i = 0...(s-1).
\label{nodeloc}
\ee
This reduces the lagrangian to the form
\be
\bL = \bL \left(g_{ab}, R_{bcde},\odel R_{bcde}, ...,  \mdel R_{bcde},
\psi, \odel \psi,..., \ldel \psi, \gammao{}'\right),
\ee
where $m = s-2 = \max (k-2,l-2)$.
The diffeomorphism invariance condition
(\ref{lagr4}) yields one more relation,
namely,
\be
\frac{\dee \bL}{\dee \gammao{}'} \Liek \gammao{}' = 0
\ee
where a
sum over the fields $\gammao{}'$ should be understood. To show that this
implies that $\bL$ has no essential dependence
on $\gammao{}'$, we proceed by introducing a local coordinate
system $x^1,..., x^n$, and viewing $\bL$ as a function of the coordinate
components of the dynamical fields and $\gammao{}'$. We then view the
components of $\gammao{}'$ as given functions of $x^{\mu}$. In this way,
we may view $\bL$ as
\be
\bL = \bL \left( g, R_{bcde},\odel R_{bcde}, ...,
\mdel R_{bcde}, \psi, \odel \psi,..., \ldel \psi, x^{\mu}\right)
\ee
i.e., we replace the dependence of $\bL$ on $\gammao{}'$ by explicit
dependence on coordinates. Condition (\ref{lagr4}) then implies that
\be
\sum_{\mu} \frac{\dee \bL}{\dee x^{\mu}} \Liek x^{\mu}= 0  .
\ee
Clearly, this equation holds for all $\k^a$ if and only if
\be
\frac{\dee \bL}{\dee x^{\mu}} = 0 .
\label{nocoord}
\ee
We therefore see that (\ref{nocoord}) implies
that any diffeomorphism invariant lagrangian
must be of the form
\be
\bL = \bL \left( g_{ab}, R_{bcde}, \odel R_{bcde}, ...,
\mdel R_{bcde}, \psi, \odel \psi, \ldel \psi\right) .
\label{finalL}
\ee
as we desired to show.
$\Box$
\end{section}

\begin{section}{The form of the symplectic potential and symplectic currents
for diffeomorphism invariant theories.}
As is well known (and as will be explicitly demonstrated in Lemma
\ref{covth} below), if we vary the dynamical fields $\phi = (g_{ab}, \psi)$,
then -- by ``integration by parts" manipulations of the terms involving
the derivatives of $\delta \phi$ -- the
first variation of the Lagrangian can always be expressed in the form
\be
\delta \bL = \bE \delta \phi + d \th
\label{delL}
\ee
with
\be
\bE \delta \phi = (\bE_g)^{ab} \delta g_{ab} + \bE_{\psi} \delta \psi
\label{Eg}
\ee
where a sum over the ``matter fields" $\psi$ is understood, and it is also
understood that for each matter field, $E_{\psi}$ has tensor indices
dual to $\psi$, and these indices
are contracted with those of $\de \psi$ in eq.(\ref{Eg}).
Here $\bE_g$ and $\bE_{\psi}$ are locally constructed out of the
dynamical fields $\phi$ and their derivatives, whereas
$\th$ is locally constructed out of $\phi$, $\delta \phi$ and their
derivatives and is linear in $\delta \phi$. The equations of motion
of the theory are then taken to be
\be
(\bE_g)^{ab} = 0 \mbox{ , and } \bE_{\psi} = 0
\ee
The $(n-1)$-form, $\th$, defined by eq.(\ref{delL}) is called the
{\em symplectic potential form} (see below). However,
although the equations of motion form, $\bE$, is uniquely
determined by eq.(\ref{delL}), this equation determines $\th$
only up to the addition of a closed (and
hence exact \cite{W2}) $(n-1)$-form. Thus, some arbitrariness is
present in the choice of $\th$. The principal result of this
section is stated in the following lemma:

\begin{lemma}
\label{covth}
Given a covariant lagrangian of the form (\ref{finalL}) one can always choose a
{\em covariant} $\th$ satisfying (\ref{delL}). Moreover,
$\th$ can be chosen to have the form
\be
\th = 2\bE_R^{bcd} \del_d \de g_{bc} + {\th}'
\label{th}
\ee
where $\th'$ is of the form
\be
\th' = \bS^{ab}(\phi) \de g_{ab} + \sum_{i=0}^{m-1}\bT_i(\phi)^{abcda_1...a_i}
\de \idel R_{abcd} + \sum_{i=0}^{l-1}\bU_i(\phi)^{a_1...a_i} \de \idel \psi
\label{thprime}
\ee
In other words, in the expression for $\th$,
the $\delta$'s can be put to the left of derivatives of the
dynamical fields everywhere except for the single
term $\bE_R^{bcd} \del_d \de g_{bc}$.
Finally, $\bE_R^{bcd}$ is given by
\be
(\bE^{bcd}_R)_{b_2...b_n} = E_R^{abcd} \eps_{a b_2...b_n}
\ee
where $E_R^{abcd} \eps_{b_1 b_2...b_n}$ is the equation of motion form that
would be obtained for $R_{abcd}$ if it were viewed as an independent
field in the Lagrangian (\ref{finalL}) rather than a quantity determined
by the metric.
\end{lemma}
{\em Proof.}  Given $\bL$ in form (\ref{finalL}) we write it as $\bL = L \eps$,
where $\eps$ is the canonical volume form on $M$ associated with
$g_{ab}$. Computing a first variation, we obtain
\bea
\de\bL&=&\eps\left(\frac{\dee L}{\dee g_{ab}}+\frac{\dee L}{\dee R_{abcd}}
 \de R_{abcd} + \frac{\dee L}{\dee \odel R_{abcd}} \de \odel R_{abcd} +...+
\frac{\dee L}{\dee \mdel R_{abcd}} \de \mdel R_{abcd} \right. \nonumber \\
& &+ \left.\frac{\dee L}{\dee \psi} \de \psi + \frac{\dee L}{\dee \odel \psi}
\de \odel \psi + \ldots + \frac{\dee L}{\ldel \psi} \de \ldel \psi \right)
+ \frac{1}{2}g^{ab} \de g_{ab} \bL
\label{delL2}
\eea
(For tensors, such as $\de R_{abcd}$, whose components
are not algebraically independent at each point, we uniquely fix the partial
derivative coefficients appearing in this
equation by requiring them to have precisely the
same tensor symmetries as the varied quantities.)
In order to obtain the desired expression for $\th$, we must suitably
rewrite eq.(\ref{delL2}) in the form (\ref{delL}). To see how this can be
done, we focus attention on a typical term,
\be
\eps \frac{\dee L}{\dee \idel R_{abcd}} \de \idel R_{abcd}
\label{typterm}
\ee
and rewrite it as,
\bea
\eps \frac{\dee L}{\dee \idel R_{abcd}} \de \idel R_{abcd}&=&
\eps \left( \frac{\dee L}{\dee\idel R_{abcd}}\odel
(\de \nstidel R_{abcd})\right)\nonumber\\
 & & + \eps \cdot \mbox{ terms proportional to } \del \de g \nonumber \\
 & = &\odel \left(\eps \frac{\dee L}{\dee \idel R_{abcd}}
      \de \nstidel R_{abcd} \right)\nonumber \\
 & &+ \odel(\eps \cdot (\mbox{terms proportional to } \de g))\nonumber\\
& &-\odel \left(\eps \frac{\dee L}{\dee \idel R_{abcd}}\right)
    \de \nstidel R_{abcd} \nonumber\\
&&+\eps \cdot (\mbox{terms proportional to } \de g) \nonumber\\
 & = & d {\bf V} - \eps \odel \left(\frac{\dee L}{\dee \idel R_{abcd}}\right)
\de \nstidel R_{abcd}  \nonumber \\
& &+ \eps \cdot (\mbox{ terms proportional to } \de g)
\eea
where the $(n-1)$-form, $\bf V$, has the form
\be
V_{b_2...b_n}=\eps_{a_1 b_2...b_n}\frac{\dee L}{\dee\tidel R_{abcd}}
\de\nstidel R_{abcd}-\eps\cdot(\mbox{terms proportional to $\de g$})
\ee
This shows that we can rewrite our original term
(\ref{typterm}) as a sum of a
similar term of lower differential order, the exact form $d{\bf V}$, and
terms proportional to $\de g$. By iterating this procedure and performing
similar manipulations on all other terms in (\ref{delL2}) containing
derivatives of variations of the curvature or matter fields, we obtain,
\be
\de \bL = \eps \left( A^{ab}_g \de g_{ab} + E^{abcd}_R \de R_{abcd} +
E_\psi \de \psi\right) + d \tilde{\th}
\label{delL3}
\ee
where the $(n-1)$-form, $\tilde{\th}$, is
covariant and has the same structure
as the right side of eq.(\ref{thprime}). Note that in eq.(\ref{delL3})
$\eps E_\psi$ are precisely the equations of motion form for the matter
fields $\psi$ and
\be
\eps E_R^{abcd} = \eps \left(\frac{\dee L}{\dee R_{abcd}}-\odel
\frac{\dee L}{\dee \odel R_{abcd}}+...+(-1)^m \mdel
\frac{\dee L}{\dee \mdel R_{abcd}}\right)
\label{ER}
\ee
would be the equations of motion form for $R_{abcd}$ {\em if} it were viewed
as an independent field.
In fact, however, $R_{abcd}$ is not an independent field, and, taking
account of the symmetries of $E_R^{abcd}$, we have
\be
E_R^{abcd} \delta R_{abcd} = 2 E_R^{abcd} \del_a \del_d \delta g_{bc}
+E_R^{abcd} {R_{abc}}^e \de g_{de}
\ee
Making this substitution and integrating twice by
parts, we obtain
{
\samepage
\bea
\de \bL & = & \eps (\tilde{A}^{bc}_g \de g_{bc}
+ 2 \del_a \del_d E^{abcd}_R \de g_{bc} + E_\psi \de \psi) \\
 & &+ d(2 \bE^{bcd} \del_d g_{bc}- 2 \del_d \bE^{bcd} \de g_{bc} + \tilde{\th})
\label{delL4}
\eea
}
where
\be
(\bE^{bcd})_{b_2...b_n}=E_R^{abcd}\eps_{a b_2...b_n}
\ee
and
\be
\tilde{A}^{bc}_g = A^{bc}_g + E_R^{pqrb} {R_{pqr}}^c.
\ee
Note that the equations of motion associated with $g_{ab}$ are thus
\be
(\bE_g)^{bc} = \eps\left(\tilde{A}^{bc}_g + 2\del_a \del_d E^{abcd}_R \right).
\label{Eg'}
\ee
Thus, we obtain
\be
\th = 2\bE_R^{bcd} \del_d \de g_{bc} + \th',
\ee
where
\be
\th' \equiv \tilde{\th} - 2 \del_d \bE_R^{bcd} \de g_{bc}.
\ee
This shows that $\th$ is manifestly covariant and of the
form claimed in the statement of the lemma. $\Box$

We comment, now, on the possible ambiguities in the choice of
$\th$ for a covariant Lagrangian. The above lemma proves that $\th$
always can be chosen to be covariant. This appears to be a very natural
requirement, and, in the following, we shall
restrict consideration to covariant choices of $\th$. The statement of
the lemma also
provides the canonical form (\ref{th}) for $\th$, which will play an important
role in our analysis below. However, this general form does not
uniquely determine $\th$, since one could add to $\th$ an exact
$(n-1)$-form which has the structure of the the right side of
eq.(\ref{thprime}). The proof of the lemma
does implicitly provide a particular algorithm
which uniquely determines a particular $\th$ from a given $\bL$, but
there does not appear to be any reason to prefer this
algorithm over other possible ones. Thus, it appears most preferable
to leave the choice of $\th$ unspecified apart from
the restriction of covariance. In fact, there exist two independent sources
of ambiguity in $\th$:
\begin{itemize}
\item As noted above, eq.(\ref{delL}) allows the freedom to alter $\th$ by
addition of an exact $(n-1)$-form
\be
\th \rightarrow \th  +  d \bY (\phi, \de \phi)
\ee
where the covariant $(n-2)$-form, $\bY$, is linear in the varied fields.
\item If we alter the lagrangian by addition of an exact $n$-form
\be
\bL \rightarrow \bL  + d \bmu ,
\label{changeL}
\ee
then the equations of motion are unaffected, so we do not alter the
dynamical content of the theory. Nevertheless, $\th$ must be shifted by
\be
\th \rightarrow \th  + \delta \bmu
\label{chth}
\ee
(If $\th$ is defined by the algorithm implicit in the proof of the above
lemma, then an additional exact term also would be added to $\th$.)
\end{itemize}
Thus, $\th$ is ambiguous up to the addition of two terms,
\be
\th \rightarrow \th  + \delta \bmu +  d \bY (\phi, \de \phi)
\label{ambigtheta}
\ee
The consequences of this ambiguity in $\th$ for the Noether current and
charge will be analyzed in the next section.

We conclude this section, by briefly reviewing the definition of the
symplectic form, $\Om$, in globally hyperbolic spacetimes
and investigating its possible ambiguities for
asymptotically flat solutions. This is of relevance here because
$\Om$ is used to define the notion of a Hamiltonian, which, in
turn, gives rise to the
notions of total energy and angular momentum. Thus,
ambiguities in $\Om$ could result in ambiguities in these notions.

Recall that the symplectic current $(n-1)$-form \cite{LW} is defined by
\be
\om (\phi, \de_1 \phi, \de_2 \phi) = \de_2 \th (\phi,\de_1 \phi)-\de_1\th
(\phi,\de_2 \phi)
\label{omega}
\ee
Let $C$ be a Cauchy surface. We take the orientation of $C$ to be
given by $\tilde{\eps}_{a_1...a_{n-1}} = n^b \eps_{ba_1...a_{n-1}}$
where $n^a$ is the future pointing normal to
$C$ and $\eps_{ba_1...a_{n-1}}$ is the positively
oriented spacetime volume form. We define
the symplectic form relative to $C$ by
\be
\Om (\phi, \de_1 \phi, \de_2 \phi) = \int_C \om (\phi, \de_1 \phi, \de_2 \phi)
\label{Omega}
\ee
(More precisely, eq.(\ref{Omega}) defines a ``pre-symplectic form" on
field configuration space. As explained in detail in \cite{LW}, the
phase space then is obtained by factoring out by
the degeneracy submanifolds of $\Om$, and $\Om$ then gives rise
to a symplectic form on phase space.) If $C$ is noncompact -- as we
assume here -- then some ``asymptotic flatness" conditions must
be imposed upon the dynamical fields, $\phi$,
(and, hence, on their variations)
in order to assure
convergence of the integral appearing in (\ref{Omega}). One normally
assumes that the metric, $g_{ab}$, approaches a flat metric, $\eta_{ab}$,
and the matter fields, $\psi$, approach zero at some suitable rate. The
precise asymptotic conditions appropriate for a given theory depend
upon the details of the theory, and, thus, must be examined on
a case-by-case basis, subject to the following general guidelines:
the asymptotic fall-off rates of the dynamical fields should be sufficiently
rapid that quantities of interest (like $\Om$, energy, and
angular momemtum) be well defined, but not so
rapid that a sufficiently wide class of solutions fails to exist. We shall not
investigate this issue further here, but will merely assume that such
suitable conditions have been imposed.

In principle,
the definition of $\Om$
depends upon the choice of $C$. However, since $d \om = 0$ whenever
$\de_1 \phi$ and $\de_2 \phi$ satisfy the linearized equations of motion
\cite{LW}, the dependence of $\Om$ on $C$ when the equations of
motion are imposed is given by an integral
of $\om$ over a timelike surface near spatial infinity. If sufficiently
strong asymptotic conditions at spatial infinity
have been imposed on the dynamical fields to assure convergence of the
integral appearing in (\ref{Omega}), then the integral of $\om$ on this
timelike surface typically will vanish, and, thus, the
definition of $\Om$ should be independent of $C$. Of course,
if this were not
automatically the case, one presumably would strengthen the
asymptotic conditions imposed upon the dynamical fields in order to make
$\Om$ be independent of $C$.

As noted above, $\th$ is ambiguous up to the terms given in
(\ref{ambigtheta}). The term involving $\bmu$ does not contribute to
$\om$ or $\Om$, so we find that the only ambiguity in $\Om$ is
\be
\Om \rightarrow \Om + \Delta \Om
\ee
with
\be
\Delta \Om =\int_\infty \de_1 \bY (\phi,\de_2 \phi )
-\de_2 \bY (\phi,\de_1 \phi).
\label{changeomega}
\ee
where the integral is taken over an $(n-2)$-sphere at spatial infinity. It
appears that the asymptotic conditions on the dynamical fields needed
to ensure the vanishing of $\Delta \Om$ typically will be weaker than
the conditions needed to ensure that $\Om$, eq.(\ref{Omega})
is well defined for a given choice of $\th$. In particular, taking account
of the difficulties in constructing a covariant $(n-2)$-form, $\bY$
out of the metric and its first variation (as well as $\eps$), we see that
for a theory in spacetime dimension $n$ in which
no matter fields are present, the asymptotic conditions
\be
g_{ab}  \sim  \eta_{ab} + o(r^{-\frac{(n-3)}{2}})
\label{decay}
\ee
together with faster fall-off conditions on derivatives of the metric,
suffice to ensure that $\Delta \Om = 0$. Thus, it does not appear that the
ambiguity in $\th$ will typically give rise to an ambiguity in the definition
of $\Om$ for suitable asymptotic conditions on the dynamical fields.
\end{section}

\begin{section}{The form of the Noether charge}

In this section, we will obtain an
expression for the general structure of the
Noether charge $(n-2)$-form, $\bQ$, for a diffeomorphism invariant
theory. We begin by reviewing the construction of
the Noether charge given in \cite{W1} (see also \cite{BCJ}).

Let $\k^a$ be any smooth vector field on the spacetime manifold, $M$,
(i.e., $\k^a$ is the infinitesimal generator of a diffeomorphism)
and let $\phi$ be any field configuration. ($\phi$ is {\em not} required,
at this stage,
to be a solution of the equations of motion). We associate to
$\k^a$ and $\phi$ a Noether current $(n-1)$-form, defined by
\be
\bJ = \th (\phi, \cal{L_\k} \phi) - \k \cd \bL
\label{Ncurrent}
\ee
where $\th (\phi, \cal{L_\k} \phi)$ denotes the expression obtained by
replacing $\delta \phi$ with $\cal{L_\k} \phi$ in the expression for $\th$,
and the ``centered dot" denotes the contraction of the vector field $\k^a$
into the first index of the differential form $\bL$.
A standard calculation (see, e.g., \cite{LW}) gives
\be
d \bJ = - \bE \cal{L_\k} \phi
\label{dJ}
\ee
which shows $\bJ$ is closed (for all $\k^a$) when the equations of motion
are satisfied. Consequently \cite{W2} there is a
$\bQ$ locally constructed from $\phi$ and $\k^a$ such that whenever
$\phi$ satisfies the equations of motion, $\bE = 0$, we have
\be
\bJ = d \bQ
\label{Q}
\ee
We refer to $\bQ$ as the Noether charge $(n-2)$-form.
Note that
for a given $\bJ$, eq.(\ref{Q}) determines $\bQ$ uniquely up to the
addition of a closed (and, hence, exact \cite{W2}) $(n-2)$-form.

\begin{proposition}
\label{Qform}
The Noether charge $(n-2)$-form can always be expressed in the form
\be
\bQ = \bW_c (\phi) \k^c + \bX^{cd} (\phi) \del_{[c} \k_{d]} +
\bY (\phi, \Liek \phi) + d \bZ (\phi, \k)
\label{Ncharge}
\ee
where $\bW_c$, $\bX^{ab}$, $\bY$, and $\bZ$ are covariant quantities
which are locally constructed from
the indicated fields and their derivatives (with $\bY$ linear
in $\Liek \phi$ and $\bZ$ linear in $\k$). This decomposition of
$\bQ$ is {\it not} unique in the sense that there are many different ways
of writing $\bQ$ in the form (\ref{Ncharge}), i.e.,
$\bW_c$, $\bX^{ab}$, $\bY$, and $\bZ$ are not uniquely determined by
$\bQ$ (see below). However, $\bX^{ab}$ may be
chosen to be
\be
(\bX^{cd})_{c_3...c_n}=-E_R^{abcd}\eps_{abc_3...c_n}
\label{XER}
\ee
where $E_R^{abcd}$ was defined by eq.(\ref{ER}), and we may choose
$\bY = \bZ = 0$.
\end{proposition}
{\em Proof.} We proceed by calculating $\bQ$ using the choice of
$\th$ given in lemma \ref{covth}, and the algorithm for calculating
$\bQ$ from $\bJ$
given in lemma 1 of \cite{W2}. For the choice of $\th$ given in lemma
\ref{covth}, we have
\be
\bJ = 2 \bE_R^{bcd} \del_d (\del_b \k_c + \del_c \k_b)
+ \th' (\phi, \Liek \phi) - \k \cdot \bL
\label{JQ}
\ee
Now the algorithm of lemma 1 of \cite{W2}
for obtaining $\bQ$ from $\bJ$ reduces the highest number of derivatives
of $\k^a$ appearing in the expression for $\bJ$
by one. Since $\th'$ is linear in the quantities
$(\de g_{ab}, \de R_{abcd}, \de \del R_{abcd}, ...)$ and does not contain
any terms involving derivatives of these quantities, it follows that
$\th' (\phi, \Liek \phi)$ is
linear in $(\k^a, \del_b \k^a)$, i.e., it has no dependence on any
derivatives of $\k^a$ higher than first. Since $\bE_R^{bcd}$ is
antisymmetric in $c$ and $d$, the term $\bE_R^{bcd} \del_d \del_c \k_b$
has no dependence on derivatives of $\k^a$. Thus, no derivatives of $\k^a$
higher than second appear in eq.(\ref{JQ}), and
only the term $\bE_R^{bcd} \del_d \del_b \k_c$
involves second derivatives
of $\k^a$. The contribution of this latter term
to $\bQ$ is readily computed, and we find that with our choice of $\th$ and
algorithm for calculating $\bQ$, we have
\be
\bQ = \bW_c (\phi) \k^c + \bX^{cd} \del_{[c} \k_{d]}
\label{Q1}
\ee
where $\bW$ is a covariant $(n-2)$-form locally constructed out of
the dynamical fields, $\phi$, and their derivatives, and where
\be
(\bX^{cd})_{c_3...c_n}=-E_R^{abcd} \eps_{abc_3...c_n}.
\label{X}
\ee

Equation (\ref{Q1}) gives the general form of $\bQ$ for our particular
algorithm for choosing $\th$ and obtaining $\bQ$ from $\bJ$.
Recall that $\th$ had two ambiguities (\ref{ambigtheta}), one arising from
the ambiguity in $\bL$ and the other from its defining
equation (\ref{delL}). Using the identity
\be
\Liek {\bf \bmu} = \k \cd d{\bf \bmu} + d(\k \cd {\bf \bmu})
\label{lieid}
\ee
we see that the ambiguity in $\th$ gives
rise to the following ambiguity in $\bJ$:
\be
\bJ \rightarrow \bJ + d(\k \cd {\bf \bmu}) + d \bY (\phi, \Liek \phi)
\label{J10}
\ee
Taking into account the additional ambiguity of addition of an exact form
to $\bQ$, we obtain the following ambiguity in $\bQ$ \cite{JKM}:
\be
\bQ \rightarrow \bQ + \k \cd {\bf \bmu} +  \bY (\phi, \Liek \phi) + d \bZ
\label{Q2}
\ee
Thus, for any choice of $\bQ$, we have
\be
\bQ=\bW_c \k^c + \bX^{cd}\del_{[c} \k_{d]} + \bY(\phi, \Liek \phi) + d\bZ.
\label{Qdecomp}
\ee
as we desired to show.
$\Box$

As stated above, the decomposition (\ref{Ncharge}) of
$\bQ$ is {\it not} unique. For example, it is clear that for any choice
of $(n-2)$-form $\bU_c (\phi)$, the quantity $d(\bU_c (\phi) \k^c)$
can be written as a sum of terms of the same form as the first three
terms on the right side of (\ref{Ncharge}), since we can write it as a sum of
a term linear in $\k^c$, a term linear in $\del_{[c} \k_{d]}$, and a term
linear in $2 \del_{(c} \k_{d)} = \Liek g_{cd}$. Thus, we can always add the
term $\bU_c (\phi) \k^c$ to $\bZ$ and make compensating changes in
$\bW$, $\bX$, and $\bY$ without affecting $\bQ$. One might be tempted
to impose additional conditions to determine the terms $\bW$, $\bX$,
$\bY$, and $\bZ$ in eq.(\ref{Ncharge}). In particular, it might appear
natural to fix the term $\bX^{ab}$ (which plays a key role in the definition
of black hole entropy below) by simply requiring it to be given by
eq.(\ref{X}) above. However, this proposal suffers from the difficulty that
a change of Lagrangian of the form (\ref{changeL}) -- which should have
no affect upon the physical content of the theory -- would, in
general produce in a change in $\bX^{ab}$. For this reason, we shall not
attempt to give unique definitions of the individual terms in
eq.(\ref{Ncharge}) but will derive the first law of black hole mechanics
and give a proposal for
defining the entropy of dynamical black holes based only
upon the general form of $\bQ$ given in proposition (\ref{Qform}).
\end{section}

\begin{section}{Examples of Lagrangians and associated
Noether currents and charges}

In this section, we shall give the symplectic potential $\th$,
the Noether current $\bJ$, and the Noether charge $\bQ$ arising
from three Lagrangians of interest. In giving these examples, we shall
simply make convenient choices of $\th$, $\bJ$, and $\bQ$, but,
of course, it should be kept
in mind that the ambiguities (\ref{ambigtheta}), (\ref{J10}), and (\ref{Q2})
remain present.

Our first example is general relativity. We have the Lagrangian $4$-form
\be
\bL_{abcd}=\frac{1}{16\pi}\eps_{abcd}R
\label{Legrav}
\ee
This yields a symplectic potential $3$-form
\be
\th_{abc}=\eps_{dabc}\frac{1}{16\pi}g^{de}g^{fh}
\left(\del_f\de g_{eh}-\del_e\de g_{fh}\right).
\label{thgr}
\ee
{}From this, we obtain the Noether current $3$-form
\be
\bJ_{abc}=\frac{1}{8\pi} \eps_{dabc}\del_e\left(\del^{[e}\xi^{d]}\right),
\ee
which yields the Noether charge $2$-form
\be
\bQ_{ab}=-\frac{1}{16 \pi} \eps_{abcd} \del^c \xi^d.
\label{Qegrav}
\ee

Our second example is 2-dimensional dilaton gravity
(in the form given in \cite{F}), with scalar field
$\phi$, coupling constant
$\lambda$, and an additional ``tachyon field" $T$.
The Lagrangian $2$-form is
\be
\bL_{ab}=\frac{1}{2}\eps_{ab} e^{\phi}\left(R+(\nabla \phi)^2-(\nabla T)^2
+\mu^2T^2 +\lambda\right)
\label{Ldil}
\ee
This yields the symplectic potential $1$-form
\be
\th_a = \eps_{ab}e^{\phi}\left[(\del^b \phi)\de \phi - (\del^b T)\de T
+ \frac{1}{2}g^{bc}(\del^d(\de g_{cd})-g^{de}\del_c(\de g_{de})-(\del^d\phi)
\de g_{cd} + (\del_c\phi) g^{de}\de g_{de})\right]
\ee
{}From this, we obtain the Noether current $1$-form
\be
\bJ_{a}=\eps_{ab}\del_c\left(e^{\phi}\del^{[c}\xi^{b]}
+2\k^{[c}\del^{b]}e^{\phi}\right),
\ee
which yields the Noether charge $0$-form (i.e., function)
\be
Q=-\frac{1}{2}\eps_{ab}\left(e^{\phi}\del^{a}\k^{b}
+2\k^{a}\del^{b}e^{\phi}\right)
\label{Qdil}
\ee

As our final example, we consider the special case of Lovelock gravity
in $n$ dimensions obtained by keeping only the terms in the Lagrangian
up to quadratic order in the curvature (see \cite{Wilt}).
The Lagrangian n-form is
\be
\bL_{a_1...a_n}=\eps_{a_1...a_n}\left(\frac{1}{16\pi}R
+ \alpha(R_{abcd}R^{abcd} - 4 R_{ab}R^{ab} +R^2)\right)
\ee
{\samepage
This yields a symplectic current $(n-1)$-form
\bea
\th_{a_1...a_{n-1}}&=&\eps_{da_1...a_{n-1}}
\left( (\frac{1}{16\pi}+2\alpha R)g^{de}g^{fh}
(\del_f\de g_{eh}-\del_e\de g_{fh}) \right.  \nonumber\\
&&+\alpha\left( -2(\del^eR)g^{df}\de g_{ef}
+4R^{de}(\del_e\de g_{fh})g^{fh}
+4R^{ef}(\del^d\de g_{ef})\right.\nonumber\\
&&\left.\left. -8R^{ef}(\del_e\de g_{fh})g^{dh}-4(\del^eR^{df})\de g_{ef}
+4R^{defh}\del_h \de g_{ef}\right)\right)
\eea
}
The corresponding Noether current $(n-1)$-form is
\be
\bJ_{a_1...a_{n-1}}=\eps_{da_1...a_{n-1}}\del_e \left( (\frac{1}{8\pi}
+4\alpha R)\del^{[e}\k^{d]}
+16\alpha(\del_f\k^{[e})R^{d]f}+4\alpha R^{edfh}\del_f\k_h\right)
\ee
which yields the Noether charge $(n-2)$-form
\be
\bQ_{a_1...a_{n-2}}=-\eps_{dea_1...a_{n-2}}\left(\frac{1}{16\pi}\del^d\k^e+
2\alpha(R\del^d\k^e+4\del^{[f}\k^{d]}{R^e}_f+R^{defh}\del_f\k_h)\right)
\label{LoveQ}
\ee
\end{section}

\begin{section}{The first law of black hole mechanics}
In this section, we will use lemma \ref{covth} and
proposition \ref{Qform} to improve upon the derivation of the
first law of black hole mechanics given in \cite{W1}. We thereby will
prove that the first law of black hole mechanics holds for nonstationary
perturbations of a black hole in an arbitrary diffeomorphism covariant
theory of gravity, without any restriction on the number of derivatives
of fields which appear in the Lagrangian.

Let $\phi$ be any solution of the equations of motion, and let $\de \phi$ be
any variation of the dynamical fields
(not necessarily satisfying the linearized equations
of motion) about $\phi$.
Let $\k^a$ be an arbitrary, fixed vector field on $M$.
We then have \cite{W1}
\bea
\de \bJ & = & \de \th (\phi, \Liek \phi) - \k \cd \de \bL \nonumber \\
& = & \de \th (\phi, \Liek \phi) -
\k \cd d \th(\phi, \de \phi)  \nonumber \\
 & = & \de \th (\phi, \Liek \phi) - \Liek \th(\phi, \de \phi) + d(\k \cd
\th(\phi, \de \phi))
\label{deJ}
\eea
where eq.(\ref{delL}) together with $\bE = 0$ was used in the second line,
and the identity (\ref{lieid}) on Lie derivatives of forms was used in
the last line. Since our choice of $\th$ is covariant,
$\Liek \th$ is the same as the variation
induced in $\th$ by the field variation $\de' \phi = \Liek \phi$.
Consequently, we have
\be
\de \th (\phi, \Liek \phi) - \Liek \th (\phi, \de \phi)
= \om (\phi, \de \phi, \Liek \phi)
\ee
where $\om$ was defined by eq.(\ref{omega}). We therefore obtain
\be
\om (\phi, \de \phi, \Liek \phi) = \de \bJ - d(\k \cd \th)
\label{deltaj}
\ee

The fundamental identity which gives rise to the first law of black hole
mechanics applies to the case where $\k^a$ is a symmetry of all of the
dynamical fields -- i.e, $\Liek \phi = 0$ -- and $\de \phi$ satisfies
the linearized equations of motion. When $\Liek \phi = 0$, the left side of
eq.(\ref{deltaj}) vanishes, and when $\de \phi$ satisfies
the linearized equations, we may replace $\de \bJ$
by $\de d \bQ = d \de \bQ$
on the right side. Thus, we obtain,
\be
d \de \bQ - d(\k \cd \th) = 0
\ee
Integrating this equation over a
hypersurface, $\Xi$, we obtain
\be
\int_{\partial \Xi} \de \bQ [\k] - \k \cd \th (\phi, \de \phi) = 0
\label{fundid}
\ee
We emphasize that the only conditions needed for the validity of
eq.(\ref{fundid}) are that
$\phi$ be a solution to the equations of motion, $\bE = 0$,
satisfying $\Liek \phi = 0$, and $\de \phi$ be a solution of
the linearized equations (not necessarily satisfying $\Liek \de \phi = 0$).

We shall be interested here in the case where
$\Xi$ is an asymptotically flat
hypersurface in an asymptotically flat spacetime. In this case,
a boundary term from an asymptotic $(n-2)$-sphere at
infinity will contribute to eq.(\ref{fundid}).
The following argument shows that this
boundary term has the natural interpretation of being the
variation of the ``conserved quantity"
canonically conjugate to the asymptotic symmetry generated by $\k^a$.

Consider a solution, $\phi$,
corresponding to an asymptotically flat, globally hyperbolic spacetime,
with Cauchy surface, $C$, having a single asymptotic
region and a compact interior.
We return to eq.(\ref{deltaj}) but no longer impose the additional
assumptions that $\Liek \phi = 0$ or that $\de \phi$ satisfy
the linearized equations of motion. We integrate eq.(\ref{deltaj}) over $C$
taking into account eq.(\ref{Omega}) and the fact that, by definition,
Hamilton's equations of motion for the dynamics generated by the time
evolution vector field $\k^a$ are
\be
\de H = \Om (\phi, \de \phi, \Liek \phi)
\label{dH}
\ee
We thereby find that {\it if} a Hamiltonian, $H$, exists for the dynamics
generated by $\k^a$, then
\bea
\de H & = & \de \int_C \bJ - \int_C d(\k \cd \th) \nonumber \\
& = & \de \int_C \bJ - \int_\infty \k \cd \th
\eea
Thus, a Hamiltonian for the dynamics
generated by $\k^a$ does exist if (and only if) we can find a (not
necessarily diffeomorphism covariant)
$(n-1)$-form, $\bB$, such that
\be
\de \int_\infty \k \cd \bB = \int_\infty \k \cd \th
\label{defB}
\ee
in which case $H$ is given by
\be
H = \int_C \bJ - \int_\infty \k \cd \bB
\label{B}
\ee
Now evaluate $H$ on solutions. We then may replace $\bJ$ by $d \bQ$,
whence $H$ becomes
\be
H = \int_\infty (\bQ - \k \cd \bB)
\label{H}
\ee
Thus, we have shown that in any theory arising from a diffeomorphism
covariant Lagrangian, the Hamiltonian -- if it exists -- always is a pure
``surface term" when evaluated ``on shell". Similarly,
for a closed universe (i.e.,
compact $C$), the Hamiltonian always vanishes ``on shell".

We now shall assume that the asymptotic conditions on the dynamical
fields have been specified in such a way that when $\k^a$ is an asymptotic
time translation, $\bB$ exists, and the surface integrals appearing in
eq.(\ref{H}) approach a finite limit at infinity. We define the {\it canonical
energy}, $\cal E$ to be the value of the Hamiltonian, i.e.,
\be
{\cal E} \equiv \int_\infty (\bQ [t] - t \cd \bB)
\label{E}
\ee
where $t^a$ is an asymptotic time translation. We then
adopt eq.(\ref{E}) as
the definition of the canonical energy associated to any asymptotically
flat region
of any solution, whether or not the spacetime is globally hyperbolic.

We illustrate this definition of canonical energy by evaluating $\cal E$ for
vacuum general relativity. We consider spacetimes which are
asymptotically flat in the sense that there exists a flat metric $\eta_{ab}$
such that in a global inertial coordinate system of $\eta_{ab}$ we have
\be
g_{\mu \nu} =  \eta_{\mu \nu} + O(1/r)
\label{gdecay}
\ee
and
\be
\frac{\dee g_{\mu \nu}}{\dee x^\alpha} = O(1/r^2).
\ee
Let $t^a$ be the asymptotic time translation $(\dee/\dee t)^a$, and
let the 2-sphere at infinity be the limit as $r \rightarrow \infty$
of the coordinate
spheres $r, t =const$. Then from
our previously calculated expression for $\bQ_{ab}$, eq.(\ref{Qegrav}),
we find that
\bea
\int_{\infty}\bQ[t] & = & - \frac{1}{16\pi} \int_{\infty}
\eps_{abcd}\del^{c}t^d \nonumber \\
& = & -\frac{1}{16\pi}\int_{\infty} dS
\left(\frac{\dee g_{tt}}{\dee r} - \frac{\dee g_{rt}}{\dee t}\right)
\label{Kom1}
\eea
Note that for a stationary spacetime
with stationary Killing field $t^a$, the first line of
eq. (\ref{Kom1}) shows that $\int_{\infty}\bQ[t]$ is precisely
one-half of the Komar mass (see, e.g., \cite{Wald}).

We now compute the contribution to ${\cal E}$ from the second
term on the right side of eq.(\ref{E}). Using eq.(\ref{thgr}), we have
\bea
\int_{\infty} t^a\th_{abc} & = & - \frac{1}{16\pi} \int_{\infty} dS
r_d g^{de}g^{fh}(\del_f\de g_{eh}-\del_e\de g_{fh}) \nonumber \\
& = & - \frac{1}{16\pi} \int_{\infty} dS g^{rr} \left(g^{tt}
(\dee_t \de g_{rt} - \dee_r \de g_{tt}) + h^{ij}(\dee_i \de h_{rj}-
\dee_r \de h_{ij})\right)\nonumber\\
&=& - \frac{1}{16\pi} \de\int_{\infty}dS
\left( (\dee_r g_{tt} - \dee_t g_{rt}) +
 r^k h^{ij}(\dee_i h_{kj} - \dee_k h_{ij})\right)
\label{bigint2}
\eea
where $r^{a} = (\dee/\dee r)^{a}$ and $h_{ij}$ is the spatial metric.
Thus, we see that eq.(\ref{defB}) holds if $\bB$ is chosen to be any
$3$-form such that asymptotically at infinity, we have
\be
t^a\bB_{abc} = - \frac{1}{16\pi} \tilde{\eps}_{bc}
\left( (\dee_r g_{tt} - \dee_t g_{rt}) +
 r^k h^{ij}(\dee_i h_{kj} - \dee_k h_{ij})\right)
\ee
where $\tilde{\eps}_{bc}$ is the volume 2-form for the sphere at infinity.
Combining this with (\ref{Kom1}) we find the that
the canonical energy, $\cal E$, for general relativity is
\bea
{\cal E} & = & \int_{\infty} \bQ[t] - t \cdot \bB \nonumber\\
&=& \frac{1}{16\pi} \int_{\infty}dS
 r^k h^{ij}(\dee_i h_{kj} - \dee_k h_{ij})
\nonumber \\
&=& M_{ADM}
\eea
where $M_{ADM}$ denotes the ADM mass. Thus, the term $t \cdot \bB$
cancels the contribution
to ${\cal E}$ from the term $\bQ$, and, in addition, provides the
term $M_{ADM}$, thereby making our definition of
${\cal E}$ in vacuum general relativity
reduce to the standard, ADM, definition
of energy. Note, however, that additional
contributions to $\cal E$ in general relativity
can occur when long range matter fields
are present; see \cite{SW} for an explicit evaluation of the contribution
to $\cal E$ for Yang-Mills fields.

When $\k^a$ is an asymptotic rotation, $\varphi^a$,
we may choose the surface at
infinity to be everywhere tangent to $\varphi^a$, in which case the pullback
of $\varphi \cd \th$ to that surface vanishes. Hence, we define the
{\it canonical angular momentum}, $\cal J$ of any asymptotic region by
\be
{\cal J} = - \int_\infty \bQ [\varphi]
\label{J}
\ee
where it is assumed that the
asymptotic conditions on the dynamical fields are such that this surface
integral approaches a well defined limit at infinity.
(The relative sign difference occuring in
the definitions (\ref{E}) and (\ref{J}) traces its origin to the Lorentz
signature of the spacetime metric. The same relative
sign difference occurs in the definitions, $E = - p_a t^a$ and
$J= + p_a \varphi^a$, of the energy and angular momentum of a particle
in special relativity.) In the axisymmetric
case in vacuum general relativity, eq.(\ref{J}) is precisely the Komar
formula for angular momentum. Thus, we see that in any theory,
the Komar-type expression
$- \int_\infty \bQ [\varphi]$ always yields the angular momentum,
but $\int_\infty \bQ [t]$ does not, in general, yield the energy.
Indeed, since the Komar and ADM masses agree for stationary solutions in
general relativity\cite{BAM}, we see from our calculation above that
$\int_\infty \bQ [t]$ yields only half of the energy in that case. It is the
presence of the ``extra"
$t \cdot \bB$ term in eq.(\ref{E})
which accounts for this well known ``factor of 2"
discrepancy in the Komar formulas for mass and angular momentum in
general relativity.

We now are ready to apply eq.(\ref{fundid}) to the case of a stationary
black hole solution with bifurcate Killing horizon. Let $\k^a$
be the killing field which vanishes on the bifurcation $(n-2)$-surface
$\Sigma$, normalized so that
\be
\k^a = t^a + \Omega_H^{(\mu)} \varphi^a_{(\mu)}
\label{k}
\ee
where $t^a$ is the stationary Killing field with unit norm at infinity, and
summation over $\mu$ is understood. (This equation picks out a family
of axial Killing fields, $\varphi^a_{(\mu)}$, acting in orthogonal planes,
and also defines the ``angular velocities of the horizon",
$\Omega_H^{(\mu)}$. No summation is required when the spacetime
dimension is less than five, and, of course, the the second term
on the right side is entirely absent in two dimensions.) Let $\Xi$ be an
asymptotically flat hypersurface having $\Sigma$ as its only
``interior boundary". Then, taking into account eq.(\ref{k}),
the definitions of ${\cal E}$ and ${\cal J}$, and the fact that $\k$ vanishes
on $\Sigma$, we obtain directly from eq.(\ref{fundid}) the result
\be
\de \int_{\Sigma} \bQ [\k] = \de {\cal E} - \Omega_H^{(\mu)} \de {\cal
J}_{(\mu)}.
\label{fl}
\ee
We now are ready to state and prove the first law of black hole mechanics
in a form which strengthens the results of \cite{W1} by establishing the
general validity of this law for nonstationary perturbations.

\begin{theorem}
\label{firstlaw}
Let $\phi$ be an asymptotically flat
stationary black hole solution with a bifurcate killing
horizon, and let $\de \phi$ be a
(not necessarily stationary), asymptotically flat solution
of the linearized equations about $\phi$. Define $S$ by,
\be
S = 2 \pi \int_{\Sigma} \bX^{cd} \eps_{cd}
\label{S}
\ee
where $\bX^{cd}$ is as given in proposition \ref{Qform}, and
the integral is taken over the bifurcation
$(n-2)$-surface, $\Sigma$, with $\eps_{cd}$ denoting the binormal
to $\Sigma$ (i.e., $\eps$ is the natural volume
element on the tangent space
perpendicular to $\Sigma$, oriented so that $\eps_{cd} T^c R^d > 0$
when $T^a$ is a future-directed timelike vector and the spacelike
vector $R^a$ points ``towards infinity"). Then we have
\be
\frac{\kap}{2 \pi} \de S = \de {\cal E} - \Omega_H^{(\mu)} \de {\cal J}_{(\mu)}
\label{1law}
\ee
where $\kap$ is the surface gravity of the black hole.
\end{theorem}
{\em Proof}: The theorem will follow from eq.(\ref{fl})
provided that we can show that
\be
\de \int_{\Sigma} \bQ [\k] = \frac{\kap}{2 \pi} \de S
\label{dS}
\ee
To evaluate the left side of this equation, we appeal to proposition
\ref{Qform} and examine the contribution of each of the four terms
individually. Since $\k$ vanishes on $\Sigma$, it is clear that the term
$\bW_c  \k^c $ contributes neither to $\bQ$ nor to its variation. Similarly,
the term $d \bZ$ clearly also makes no contribution to the left side
of eq.(\ref{dS}). Since $\Liek \phi = 0$, the term $\bY$ vanishes in the
stationary background, and its first variation is given by
\bea
\de \bY (\phi, \Liek \phi) & = & \bY (\phi, \Liek \de \phi) \nonumber \\
& = & \Liek \bY (\phi, \de \phi) \nonumber \\
& = & \k \cd d \bY + d(\k \cd \bY)
\eea
where the Lie derivative identity (\ref{lieid}) was used in the last line. It
follows immediately that the term $\bY$ also makes no contribution to
the left side of eq.(\ref{dS}). Thus, we have
\be
\de \int_{\Sigma} \bQ [\k] = \de \int_{\Sigma} \bX^{cd} (\phi) \del_{[c}
\k_{d]}
\label{dQ}
\ee
Now, in the stationary background, we have, on $\Sigma$
\be
\del_{c} \k_{d} = \kap \eps_{cd}
\ee
Furthermore, since $\k^a = 0$ on $\Sigma$, and $\de \k^a = 0$
everywhere, we have
\be
\de \del_{c} \k^{d} = 0
\ee
on $\Sigma$. Consider, now, the variation, $\de {\eps_c}^d$ of the
binormal, ${\eps_c}^d$, with an index raised. Clearly,
$s^c \de {\eps_c}^d = 0$ for all $s^c$ tangent to $\Sigma$, so
$\de {\eps_c}^d$ has no ``tangential-tangential" piece. However, since
${\eps_c}^d {\eps_d}^c$ does not vary as the metric is changed, it
follows that $g_{a[c} \de {\eps_{d]}}^a$ has no ``normal-normal'' piece
with respect to the background metric. Thus, writing
\be
w_{cd} = \del_{[c} \k_{d]} - \kap \eps_{cd}
\ee
we have that $w_{cd}$ vanishes in the stationary background, and
\bea
\de w_{cd}&=&\de\left(g_{a[d}(\del_{c]}\k^{a}
-\kap{\eps_{c]}}^a)\right)\nonumber \\
& = & - \kap g_{a[d} \de {\eps_{c]}}^a
\eea
so that $\de w_{cd}$ has only a ``normal-tangential" piece with respect
to the background metric. Thus, substituting in eq.(\ref{dQ}), we find,
\bea
\de \int_{\Sigma} \bQ [\k] & = & \de \int_{\Sigma} \bX^{cd} (\phi)
[\kap \eps_{cd} + w_{cd}] \nonumber \\
& = & \frac{\kap}{2 \pi} \de S + \int_{\Sigma} \bX^{cd} \de w_{cd}
\label{Xw}
\eea
Finally, we note that since $\Liek\phi = 0$, we have $\Liek \bX^{cd} = 0$,
and, hence, by lemma 2.3 of \cite{KW}, at each point of $\Sigma$, $\bX^{cd}$
must be invariant under ``reflections" about $\Sigma$, i.e.,
$\bX^{cd}$ must be invariant under the map of the tangent space which
reverses the normal directions to $\Sigma$ but keeps the tangential
directions unchanged. On the other hand, since $\de w_{cd}$ is purely
``normal-tangential", it reverses sign under reflections about $\Sigma$.
However, the pull-back of $\bX^{cd} \de w_{cd}$ to $\Sigma$ is purely
tangential, and, hence, invariant under reflections. Consequently, the
pull-back of $\bX^{cd} \de w_{cd}$ to $\Sigma$ must vanish, so the second
term on the right side of eq.(\ref{Xw}) does not contribute.
$\Box$

It should be noted that in the above discussion, $\Sigma$ was explicitly
chosen to be the bifurcation surface of a bifurcate Killing horizon. However,
as pointed out in \cite{JKM}, for a stationary black hole with bifurcate
horizon,
the integral of $\bQ$ is independent of the choice of
horizon cross-section. Namely, the difference between the integrals of
$\bQ$ over cross-sections $\Sigma$ and $\Sigma'$ is given by an integral
of $\bJ$ over the intervening portion of the horizon. However, by
eq.(\ref{Ncurrent}), the pullback of $\bJ$ to the horizon vanishes, since
${\cal L}_\k \phi = 0$ and the pullback of  $\k \cd \bL$ vanishes since
$\k^a$ is tangent to the horizon.

Furthermore, if we define the entropy, $S$, for an arbitrary
horizon cross-section, $\Sigma'$, of a stationary black hole by
\be
S[\Sigma'] = 2 \pi \int_{\Sigma'} \bX^{cd} \eps'_{cd}
\label{S'}
\ee
where $\eps'_{cd}$ denotes the binormal to $\Sigma'$, then $S$ also is
independent of the choice of $\Sigma'$ \cite{JKM}. To
prove this, we note that since $\bX^{cd}$ is invariant under the
one parameter group of isometries,
$\chi_t$, generated by $\k^a$, it follows immediately that
$S[\chi_t(\Sigma')] = S[\Sigma']$. However, as $t \rightarrow - \infty$,
$\chi_t(\Sigma')$ continuously approaches
the bifurcation surface, $\Sigma$, and (since $\bX^{cd}$
is smooth) we thus obtain $S[\Sigma'] = S[\Sigma]$, as we desired to
show. It follows immediately that for stationary perturbations, the first
law of black hole mechanics (\ref{1law}) holds with $S$ taken to be the
entropy of an arbitrary cross-section of the horizon. However, when
nonstationary perturbations are considered, it is essential for the validity
of eq.(\ref{1law}) that $S$ be evaluated on the bifurcation surface,
$\Sigma$.

As emphasized at the end of the section 4, the decomposition of
$\bQ$ given by eq.(\ref{Ncharge}) does not uniquely determine $\bX^{cd}$.
Nevertheless, theorem \ref{firstlaw} and its proof show that all of the
different possible choices of $\bX^{cd}$ yield the same value of the
entropy, $S$, for a stationary black hole. Furthermore, even for
nonstationary perturbations, the first variation,
$\de S$, of $S$ on $\Sigma$ is independent of the choice of $\bX^{cd}$.
However, for nonstationary perturbations,
$\de S$ will, in general, depend upon the choice of $\bX^{cd}$ when
evaluated on an arbitrary cross-section, $\Sigma'$, of the horizon,
and the dependence of $S$ upon the choice of $\bX^{cd}$
becomes even more severe if we attempt to generalize the
notion of entropy to an arbitrary cross-section of a
nonstationary black hole
via eq.(\ref{S'}).
We turn, now to an analysis of the definition of entropy for
nonstationary black holes.
\end{section}

\begin{section}{A prescription for dynamical black hole entropy}
In this section we will suggest a definition of the entropy, $S_{dyn}$, for a
``dynamical" (i.e., nonstationary) black hole.
We seek a formula of the general type
\be
S_{dyn} [{\cal C}] = \int_{\cal C} \bXt^{cd}(\phi) \eps_{cd}
\label{Sdyn1}
\ee
where $\cal C$ is an arbitrary cross-section of the event horizon of a
dynamical black hole, and
$\bXt^{cd}$ is a diffeomorphism covariant $(n-2)$-form
locally constructed out of the dynamical fields, $\phi$, and their
derivatives by an algorithm whose sole input is the Lagrangian, $\bL$.
There are four basic criteria which our definition of
$S_{dyn}$ must satisfy:
\begin{enumerate}
\item For an arbitrary cross-section, $\Sigma'$,
of a stationary black hole, we must have
\be
S_{dyn} [\Sigma'] = S[\Sigma'] = 2 \pi \int_{\Sigma'} \bX^{cd} \eps'_{cd}
\ee
(see eq.(\ref{S'}) above).
\item For an arbitrary (nonstationary) perturbation of a stationary black
hole, on the bifurcation surface, $\Sigma$, we must have
\be
\delta S_{dyn} [\Sigma] = \delta S =
2 \pi \delta \int_{\Sigma} \bX^{cd} \eps_{cd}
\ee
(see eq.(\ref{S}) above).
\item If we alter the Lagrangian by addition of an exact $n$-form
\be
\bL \rightarrow \bL  + d \bmu
\ee
then the definition of $S_{dyn}$ should not change, since there is no
change in the dynamical content of the theory.
\item At least for an appropriate class of theories, $S_{dyn}$ should obey
a ``second law", i.e., $S_{dyn}$ should be a
non-decreasing quantity when evaluated on successively ``later"
cross-sections of the horizon of a dynamical black hole.
\end{enumerate}

The last of these criteria is by far the most interesting and important.
Unfortunately, it also is the most difficult to analyze in a general theory
of gravity for at least the following two reasons: First, it seems clear that,
unlike the ``first law", any proof of the second law would need
to make detailed use of the equations of motion of the theory.
Second, it seems clear that the ``second law" should
hold only for the case of theories which satisfy certain
physically reasonable criteria, likely examples of which are
the existence of a well posed
initial value formulation, cosmic censorship,
and the property of having positive total energy.
For example, even for general relativity, the second law can fail if matter
is present which fails to satisfy the weak energy condition.
However, it is far from clear as to precisely what conditions should be
imposed upon a theory for
the validity of the second law to hold, and -- even if these conditions
were known -- it undoubtedly would be highly
nontrivial to determine whether a given theory satisfied them.

Despite these two difficulties, there are some hints that
it may be possible to prove some general results pertaining to the second
law. In particular, we saw in the previous section that
the entropy, $S$, of a stationary black hole is just its Noether charge with
respect to the horizon Killing field, $\k^a$. Thus, the change in entropy
between cross-sections ${\cal C}$ and ${\cal C}'$
of a stationary black hole is given by the flux of the corresponding
Noether current
through the horizon between ${\cal C}$ and ${\cal C}'$. For a stationary
black hole, this flux, of course, vanishes. However, if $S_{dyn}$ could
similarly be identified as the Noether charge of an appropriate
vector field, one might be able to
establish a relationship between the ``second law" and positive energy
(i.e., positive net Noether flux) properties of the theory.
Another suggestive fact is that
the quantity $\bX^{cd}$ which plays a key role in the definition of
entropy for stationary black holes can be chosen to be very simply
related to $E_R^{abcd}$ (see eq.(\ref{XER}) above), and $E_R^{abcd}$,
in turn, is a term in the equations of motion (see eq.(\ref{Eg'}) above).
Thus, there is a hint that it may be possible to define $S_{dyn}$ in such
a way that its dynamical properties may be directly related to the
equations of motion of the theory. Unfortunately, we have not, as yet,
succeeded in developing either of
these hints into any results regarding proposed
definitions of $S_{dyn}$. Thus, for the remainder of this section, we shall
not consider criterion (4) further, and will merely seek a definition of
$S_{dyn}$ which satisfies conditions (1)-(3).

An obvious first try at defining
$S_{dyn}$ via an equation of the form (\ref{Sdyn1}) would be to simply
set $\bXt^{cd} = \bX^{cd}$, with $\bX^{cd}$ given
by the decomposition (\ref{Qdecomp}) of $\bQ$.
However, we already emphasized above that this decomposition is not
unique. Although -- as discussed at the end of the previous section --
this ambiguity does not affect the evaluation of $S$ on an arbitary
cross-section of a stationary black hole or the evaluation of
$\delta S$ on the bifurcation surface of a stationary black hole,
this ambiguity in $\bX^{cd}$ is of importance for a dynamical black hole.

An obvious try at circumventing this difficulty would be to continue to
set $\bXt^{cd} = \bX^{cd}$ and simply fix $\bX^{cd}$
by some definite algorithm. In particular, the choice
\be
\bX^{cd}_{a_3...a_n} = -E_R^{abcd} \eps_{aba_3...a_n}.
\label{try2}
\ee
(see eq.(\ref{XER}) above) appears to be particularly simple and
natural. This proposed definition of $S_{dyn}$ clearly satisfies
conditions (1) and (2) above. However, it is not difficult to verify
that it fails \cite{JM2} to satisfy condition (3): By adding an exact form to
$\bL$ which has suitable dependence upon the curvature,
we can alter $E_R^{abcd}$ in such a way as to produce nonvanishing
changes in $S_{dyn}$ for nonstationary black holes.
We feel that it is unlikely that any other simple algorithm for fixing
$\bX^{cd}$ for a given $\bL$ will fare any better in this regard.

Thus, it is a nontrivial challenge to find {\em any} prescription for
$S_{dyn}$ of the form (\ref{Sdyn1}) which satisfies conditions (1)-(3).
We now shall demonstrate that such a prescription does exist.
The basic idea will be to construct new dynamical fields relative to a
cross section $\cal C$, which make $\cal C$ ``look like" a bifurcation
surface of a stationary black hole. We then shall define $S_{dyn}[{\cal C}]$
to be the entropy of this stationary black hole.
Before giving a precise statement of our prescription, we give the
following two definitions:

\begin{defn}: Let ${\cal C}$ be a $(n-2)$-dimensional spacelike
surface in an $n$-dimensional spacetime, and let
${M^{a_1...a_k}}_{b_1...b_l}$ be a (spacetime) tensor field defined
on ${\cal C}$. Then ${M^{a_1...a_k}}_{b_1...b_l}$ is said to be
{\em boost invariant} on ${\cal C}$ if, for each $p \in {\cal C}$,
${M^{a_1...a_k}}_{b_1...b_l}$ is invariant under Lorentz boosts
in the tangent space at $p$ in the
$2$-dimensional timelike plane orthogonal to ${\cal C}$.
\end{defn}

The following simple criterion can be used to check if a tensor field
${M^{a_1...a_k}}_{b_1...b_l}$ is boost invariant on ${\cal C}$. At each point
$p \in {\cal C}$, choose a null tetrad with null vectors $\l^a$ and $n^a$
orthogonal to ${\cal C}$, and spacelike vectors $s_{\mu}^a$ tangent to
${\cal C}$. Expand ${M^{a_1...a_k}}_{b_1...b_l}$ in this basis. Then it is
easy to verify that ${M^{a_1...a_k}}_{b_1...b_l}$ is boost invariant if and
only if its basis expansion is ``balanced" with respect to $\l^a$ and $n^a$,
i.e., if the basis expansion coefficients are nonvanishing only for terms
involving equal numbers of $\l^a$'s and $n^a$'s. This motivates the
following definition.

\begin{defn}: Let ${\cal C}$ be a $(n-2)$-dimensional spacelike
surface in an $n$-dimensional spacetime, and let
${M^{a_1...a_k}}_{b_1...b_l}$ be a (spacetime) tensor field defined
on ${\cal C}$. We define the {\em boost invariant part} of
${M^{a_1...a_k}}_{b_1...b_l}$ to be the tensor field on ${\cal C}$ obtained
by keeping only the terms which are balanced with respect to $\l^a$
and $n^a$ in a null tetrad basis expansion.
\end{defn}

It is easily seen that the boost invariant part of
${M^{a_1...a_k}}_{b_1...b_l}$ does not depend upon the choice of null
tetrad appearing in the definition.

Note that the spacetime metric, $g_{ab}$, on ${\cal C}$ is automatically
boost invariant. However, the curvature of $g_{ab}$ and its
derivatives need not be. Nevertheless, we may define a notion of
the boost invariant part (up to order $q$),
${g}^{I_q}_{ab}$, of the spacetime metric
in a neighborhood of ${\cal C}$. The curvature of
${g}^{I_q}_{ab}$ and its covariant derivatives up to order $(q-2)$ then
will automatically be boost invariant on ${\cal C}$. This construction
of ${g}^{I_q}_{ab}$ will lead directly to a proposal for defining $S_{dyn}$.

To define ${g}^{I_q}_{ab}$,
it is convenient to introduce a coordinate system in a
neighborhood of ${\cal C}$ as follows \cite{KW}. Define a null tetrad
$l^a, n^a, s_{\mu}^a$ on ${\cal C}$ as above, with $l^a n_a = -1$.
Let $\cal O$ be any neighborhood
of $\cal C$ sufficiently small that
each point $x \in {\cal O}$ lies on a unique geodesic orthogonal to $\cal C$.
Given $x \in {\cal O}$ we find the point
$p \in {\cal C}$ and the geodesic tangent
$v^a$ in the $(n-2)$-plane normal to $\cal C$ such that $x$ lies at unit
affine parameter along the geodesic determined by $p$ and $v^a$.
We assign the coordinates $(U,V,s_1,...s_{n-2})$ to $x \in {\cal O}$
by taking $(U,V)$ to be the components of $v^a$ along $l^a$ and
$n^a$, respectively, and taking $s_i$ to be (arbitrarily chosen)
coordinates of $p$ on $\cal C$. We denote by $\partial_a$ the flat
derivative operator associated with these coordinates. Note that a
change in tetrad, $l^a \rightarrow \alpha l^a$,
$n^a \rightarrow \alpha^{-1} n^a$, at $p \in {\cal C}$ (corresponding to a
Lorentz boost
in the tangent space in the plane orthogonal to ${\cal C}$)
induces the
linear change in coordinates, $U \rightarrow \alpha^{-1} U$,
$V \rightarrow \alpha V$, $s_i \rightarrow s_i$. Since linearly
related coordinate systems define the same ``ordinary derivative
operator", it follows that $\partial_a$ does not depend upon the
choice of $l^a$ and $n^a$, and so is invariant under the action of Lorentz
boosts in the plane orthogonal to ${\cal C}$.

Now consider the first $q$ terms in the Taylor series expansion of $g_{ab}$
around $\cal C$ in $U$ and $V$
\be
\left. g^{(q)}_{ab}(x^{\mu}) = \sum_{n,m = 0}^{q} \frac{U^m V^n}{m!n!}
\sum_{\alpha \beta}
\frac{\dee^{m+n} g_{\alpha \beta}}{\dee^m U \dee^n V}(s_i)\right|_{U=V=0}
(dx^{\alpha})_a (dx^{\beta})_b.
\label{gexp}
\ee
The coefficients appearing in this expansion are just components of
the tensors $\partial_{c_1}...\partial_{c_r} g_{ab}$ on ${\cal C}$,
namely,
\be
\left.\sum_{\alpha \beta}
\frac{\dee^{m+n} g_{\alpha \beta}}{\dee^m U \dee^n V}(s_i)\right|_{U=V=0}
(dx^{\alpha})_a (dx^{\beta})_b =
l^{c_1}...l^{c_m} n^{c_1}...n^{c_n} \partial_{c_1}...\partial_{c_{m+n}} g_{ab}
\ee
We define ${g}^{I_q}_{ab}$ by replacing each tensor,
$\partial_{c_1}...\partial_{c_r} g_{ab}$, appearing in the expansion of
$g^{(q)}_{ab}$ by its boost invariant part. In other words, we alter
$g_{ab}$ by extracting the boost invariant part of the coefficients of
the first $q$ terms of its Taylor expansion in $U$ and $V$.

The nature of ${g}^{I_q}_{ab}$ can be best elucidated in the case where
$g_{ab}$ is analytic, in which case we may set $q = \infty$
and write ${g}^{I}_{ab}$ for ${g}^{I_\infty}_{ab}$. It then follows
that the vector field
\be
\k^a = U (\frac{\dee}{\dee U})^a - V (\frac{\dee}{\dee V})^a
\label{Kvf}
\ee
(which induces Lorentz boosts of the coordinates)
is a Killing field of the metric ${g}^{I}_{ab}$, that is
\be
\Liekt {g}^{I}_{ab} = 0.
\ee
Furthermore, $\k^a$ vanishes on ${\cal C}$. Thus, our construction of
${g}^{I}_{ab}$ has, in effect, created a new spacetime (which is {\em not}
necessarily a solution of the field equations) in which ${\cal C}$ is the
bifurcation surface of a bifurcate Killing horizon.

In an exactly
similar manner, we define the boost invariant part, $\psi^{I_q}$ of the
matter fields $\psi$ (up to order $q$) by
extracting the boost invariant part of the coefficients of
the first $q$ terms of the Taylor expansion
of $\psi$ in $U$ and $V$ about $\cal C$.
It then follows that $\k$ also Lie derives $\psi^{I_q}$ up to order $q$.

Our proposal for defining $S_{dyn}$ is the following: Choose $q$ to be
larger than the highest derivative of any dynamical field appearing in
the decomposition of $\bQ$ given in proposition \ref{Qform}.
Given a cross-section, $\cal C$, of the horizon of a black hole, we replace
${g}_{ab}$ by ${g}^{I_q}_{ab}$ and $\psi$ by $\psi^{I_q}$ in a neighborhood
of ${\cal C}$. Define $\tilde{\bQ} [\k]$ on
${\cal C}$ to be the Noether charge
$(n-2)$-form of the dynamical fields
$\phi^{I_q} = (g^{I_q}_{ab}, \psi^{I_q})$ for the
vector field $\k^a$ defined by eq.(\ref{Kvf}) above. Define $S_{dyn}$
at ``time" ${\cal C}$ by
\be
S_{dyn} [{\cal C}] = 2 \pi \int_{{\cal C}} \tilde{\bQ} [\k]
\label{tQ}
\ee
Equivalently, by proposition \ref{Qform} we have
\be
S_{dyn} [{\cal C}] = 2 \pi \int_{\cal C} \bXt^{cd} \eps_{cd}
\label{S3}
\ee
where
\be
\bXt^{cd} (\phi) \equiv \bX^{cd} (\phi^{I_q})
\ee
Equation (\ref{S3}) shows that $S_{dyn}$ is of the desired
general form (\ref{Sdyn1}), and
the equivalence of eqs.(\ref{tQ}) and (\ref{S3}) shows that the
right side of (\ref{S3}) does not depend upon the choice of $\bX^{cd}$
in the decomposition of proposition \ref{Qform}.
Note, incidentally, that since $\bX^{cd}$ is a nonlinear function
of the dynamical fields $\phi$, the tensor field $\bXt^{cd}$ is
{\em not} necessarily equal to the
``boost invariant part" of the tensor field $\bX^{cd}(\phi)$.
(Use of the boost invariant part of $\bX^{ab}(\phi)$ would not yield
a satisfactory prescription for $S_{dyn}$ since it would, in general, fail to
satisfy condition (3) above.)
More generally, for a nonlinear
tensor function $\beta$ of the dynamical fields $\phi$,
we have,
in general, $[\beta (\phi)]^{I_q} \neq \beta (\phi^{I_q})$. On the other
hand, if $\beta$ is linear in $\phi$,
then $[\beta (\phi)]^{I_q} = \beta (\phi^{I_q})$.

We now may verify that our definition of $S_{dyn}$
satisfies conditions (1)-(3) above. First, if ${\cal C}$ is taken to be the
bifurcation surface, $\Sigma$, of a stationary black hole, then
$\phi^{I_q} = \phi$, so, clearly, $S_{dyn} [\Sigma] = S [\Sigma]$. On the
other hand, if $\Sigma'$ is an arbitrary cross-section of a stationary
black hole, then since our prescription for defining $S_{dyn}$
is a ``local, geometrical" one, by isometry invariance we clearly have
$S_{dyn} [\chi_t (\Sigma')] = S_{dyn} [\Sigma']$. But it also is clear that
our prescription for defining $S_{dyn}$ is such that $S_{dyn} [{\cal C}]$
varies continuously with ${\cal C}$. From these facts, it follows
immediately by the same argument as given below eq.(\ref{S'}) that
$S_{dyn} [\Sigma'] = S_{dyn} [\Sigma]$. Thus, we have
\be
S_{dyn} [\Sigma'] = S_{dyn} [\Sigma] = S [\Sigma] = S [\Sigma']
\ee
i.e., condition (1) is satisfied.

To verify that condition (2) holds, we note that since
we have $\phi^{I_q} = \phi$
on the bifurcation surface, $\Sigma$, of a stationary black hole, and since
$\delta [\bX^{cd} \eps_{cd}]$ clearly is linear in $\delta \phi$, it follows
that $\delta [(\bXt^{cd} - \bX^{cd}) \eps_{cd}]$ has no boost invariant
part. However, this immediately implies that the pullback of this
differential form to $\Sigma$ vanishes, from which it follows that
$\delta S_{dyn} [\Sigma] = \delta S [\Sigma]$, as desired.

Finally, the complete ambiguity in $\bQ$ (including that arising from
the change in Lagrangian $\bL \rightarrow \bL  + d \bmu$) is given by
eq.(\ref{Q2}). It is manifest that none of these ambiguous terms can
contribute to $\int_{{\cal C}} \tilde{\bQ} [\k]$. Consequently, we see
from eq.(\ref{tQ}) that condition (3) holds.

Thus, we have proven the existence of a
definition of $S_{dyn}$ which satisfies conditions (1)-(3). These conditions
do {\em not} uniquely determine $S_{dyn}$. Nevertheless, we have
been unable to come up with any
``natural" alternative definitions of $S_{dyn}$. Thus, we believe that
our definition of $S_{dyn}$ is a serious candidate for the definition
of the entropy of a nonstationary black hole in a general theory of
gravity.

We conclude this section by evaluating $S_{dyn}$ for the three theories
considered in section 5.  Consider, first, vacuum general relativity.  Let
${\cal C}$
be an arbitrary cross-section of a black hole, let $\eps_{ab}$ be the
binormal to ${\cal C}$, and let $\tilde{\eps}_{ab}$ denote
the volume element on ${\cal C}$. Comparing eqs.(\ref{Ncharge})
and (\ref{Qegrav}), we see that the $2$-form $\bX^{cd}$ is given by simply
\be
(X^{cd})_{ab} = - \frac{1}{16 \pi} {\eps_{ab}}^{cd}
\ee
Since $\bX^{cd}$ does not depend upon any derivatives of $g_{ab}$,
it is clear that it is unaffected when $g_{ab}$ is replaced by its boost
invariant part. Thus, we obtain,
\bea
S_{dyn}[{\cal C}] & = & -\frac{1}{8}
\int_{\cal C}{\eps_{ab}}^{cd} \eps_{cd} \nonumber \\
& = & \frac{1}{4} \int_{\cal C} \tilde{\eps}_{ab} \nonumber \\
& = & \frac{\mbox{Area}[{\cal C}]}{4}
\label{A}
\eea
in agreement with the usual formula for the entropy of a dynamical
black hole in general relativity.
By the area theorem, this definition of $S_{dyn}$
satisfies the ``second law" (assuming that the cosmic
censor hypothesis is valid). Note that if we add to the Lagrangian
``matter terms" which have no explicit dependence upon the curvature,
then $\bX^{cd}$ does not change (see eq. (\ref{XER})),
so eq.(\ref{A}) also holds for general relativity with matter present,
provided only that the matter
does not have an explicit coupling to the curvature in the Lagrangian.

The calculation of $S_{dyn}$ for dilaton gravity with
Lagrangian (\ref{Ldil}) in two spacetime dimensions proceeds similarly.
We see from eq. (\ref{Qdil}) that the $0$-form $X^{cd}$ is given by
\be
X^{cd} = - \frac{1}{2}e^{\phi} \eps^{cd}
\ee
Again $\bX^{cd}$ does not depend upon any derivatives of the dynamical
fields, and is unchanged when they are replaced by their boost
invariant parts. In this case, a cross-section, $\cal C$, of the horizon is a
point, and we obtain
\be
S_{dyn}[{\cal C}] = \left. 2 \pi e^{\phi} \right|_{\cal C}
\ee
It is known that this definition of $S_{dyn}$ also satisfies the second law
\cite{F}.

Lovelock gravity provides a more interesting illustration of our
prescription, since it can be seen from eq.(\ref{LoveQ})
that $\bX^{cd}$ contains terms involving the curvature,
\be
(\bX^{cd})_{a_1...a_{n-2}} = -{\eps^{cd}}_{a_1...a_{n-2}}\left(
\frac{1}{16\pi}+ 2\alpha R \right)-
8\alpha{\eps^{[d}}_{fa_1...a_{n-2}} R^{c]f}-
2\alpha\eps_{fha_1...a_{n-2}}R^{cdfh}
\label{LoveX}
\ee
and the replacement of the metric by its boost invariant part will have
a nontrivial effect. Indeed, since, after this replacement is made,
both extrinsic curvatures of $\cal C$ embedded in $M$ vanish,
we see (using a ``Gauss-Codazzi" equation -- see, e.g., \cite{Wald})
that the curvature of the boost invariant part of the metric satisfies,
\be
{^{(n-2)}\! R} = R - 2 t^{ab}R_{ab} + t^{ac}t^{bd}R_{abcd}
\label{GC}
\ee
where $t_{ab} = - n_an_b + r_ar_b$ is the metric for the
subspace orthogonal to $\cal C$ (spanned by the unit timelike and spacelike
normals, $n^a$ and $r^a$ respectively) and $^{(n-2)}\!R$ is the scalar
curvature
of $\cal C$. From eqs.(\ref{LoveX}) and (\ref{GC}), we obtain
\be
\eps_{cd} \tilde{\bX}^{cd}{}_{a_1...a_n}
=\left(\frac{1}{8\pi}+4\alpha\, {{}^{(n-2)}\! R}[g^{I_q}]\right)
\tilde{\eps}_{a_1...a_{n-2}}
\ee
where $^{(n-2)}R[g^{I_q}]$ is the $(n-2)$-scalar curvature of
$\cal C$ computed with the boost invariant part of the
metric and $\tilde{\eps}_{a_1...a_{n-2}}$ is the volume form $\cal C$. However,
we clearly
have $^{(n-2)}R[g^{I_q}] = ^{(n-2)}\! R[g]$. Hence, we obtain,
\be
S_{dyn} = \frac{1}{4} \mbox{Area}[{\cal C}]
+ 8\pi\alpha\int_{\cal C} {^{(n-2)}R}
\ee
Note that this formula differs from what would be obtained from
simply substituting the expression (\ref{LoveX}) into eq.(\ref{S}). It is not
known whether this definition of $S_{dyn}$ satisfies the second law.
\end{section}

\begin{section}*{Acknowlegements}
This research was supported in part by NSF Grant PHY-9220644 to the University
of Chicago.
\end{section}

\begin{section}*{Appendix: Applications to Theories with a Nondynamical
\linebreak Metric}
In the body of this paper, we have considered theories which are
diffeomorphism covariant in the sense of eq.(\ref{lagr2}). It
was seen in section 2 that this condition implies the absence of
``nondynamical fields" in the Lagrangian. In particular, the diffeomorphism
covariance condition excludes the case of theories with a nondynamical
metric, such as theories of fields in flat spacetime. Nevertheless, a number
of formulas and results derived in the body of this paper continue to
hold for theories with a Lagrangian locally constructed out of a
metric, $g_{ab}$, and matter fields $\psi$, of the form
(\ref{newl}), i.e., for the Lagrangian
\be
\bL = \bL \left(g_{ab}, \del_{a_1}R_{bcde},...,
\del_{(a_1}...\del_{a_m)}R_{bcde},
\psi, \del_{a_1}\psi, \del_{(a_1}...\del_{a_l)} \psi\right)
\label{new1'}
\ee
but where the metric, $g_{ab}$, is now treated as a fixed, nondynamical
entity, so that, in particular, the equations of motion, $\bE_g = 0$,
no longer are imposed. The
purpose of this appendix is to present simple, unified derivations
of some formulas and results (most of which are ``well known")
for such theories with a nondynamical metric.

In a theory with Lagrangian of the form (\ref{new1'}) but
with nondynamical metric, we define the {\em stress-energy tensor},
$T^{ab} = T^{(ab)}$, of the matter fields by,
\be
T^{ab} \eps = 2 (\bE_g)^{ab}
\label{defT}
\ee
For each vector field, $\k^a$, we
again define the Noether current, $\bJ$, by eq.(\ref{Ncurrent}) above.
However, the (matter) equations of motion no longer imply that $\bJ$
is closed. Indeed, by eq.(\ref{dJ}), we see that when $\bE_\psi = 0$,
we have
\bea
d \bJ & = & - (\bE_g)^{ab} {\cal L}_\k g_{ab} \nonumber \\
& = & - T^{ab} \del_{(a} \k_{b)} \eps \nonumber \\
& = & - \del_a [T^{ab} \k_b] \eps + \k_b \del_a [T^{ab}] \eps \nonumber \\
& = & - d(k \cdot \eps) + \k_b \del_a [T^{ab}] \eps
\label{cons}
\eea
where
\be
k^a \equiv T^{ab} \k_b
\ee
By inspection of eq.(\ref{cons}), we see that the $n$-form
$\k_b \del_a [T^{ab}] \eps$ is exact for all $\k^a$. However, since
$\k^a$ is arbitrary, this is impossible unless
\be
\del_a T^{ab} = 0
\ee
which shows that
the stress-energy tensor is covariantly conserved whenever the
matter equations of motion hold. Note that eq.(\ref{cons}) then yields
simply
\be
d (\bJ + k \cdot \eps) = 0
\ee
from which it follows immediately \cite{W2} that
\be
\bJ + k \cdot \eps = d \bK
\label{K}
\ee
where $\bK$ is locally constructed out of $g_{ab}$, $\psi$, $\k^a$, and
their derivatives. In other words, we have shown that, apart from a
``surface term", the Noether current is equivalent to the
stress-energy current $- T^{ab} \k_b$.

It is important to note that the equations of motion for $g_{ab}$ were
not used anywhere in the derivation of eqs.(\ref{deJ})-(\ref{deltaj}) or
eqs.(\ref{dH})-(\ref{B}). Thus, these equations remain valid in the case
of a theory with a nondynamical metric. In particular, if a Hamiltonian
exists for a time translation vector field, $t^a$, on a globally
hyperbolic, asymptotically flat spacetime, then it is natural to
define the canonical energy at ``time" $C$ by
\be
{\cal E} = \int_C \bJ - \int_\infty t \cd \bB
\label{E5}
\ee
(see eqs.(\ref{B}) and (\ref{E}) above). In other words, apart from the
possible ``surface term" $\int_\infty t \cd \bB$ (which
vanishes in most of the commonly considered theories of matter fields
in a background spacetime), the
canonical energy is simply the integral of the Noether current, $\bJ$, over
a Cauchy surface. However, since we no longer have $\bJ = d \bQ$, this
volume integral no longer can be converted into a surface integral.
In particular,
${\cal E}$ depends upon the choice of $t^a$ in the interior of the spacetime,
not just upon its asymptotic value at infinity. Note also that since $\bJ$
need not be closed (see eq.(\ref{cons})),
${\cal E}$ need not be conserved, i.e., independent of $C$.

Using eq.(\ref{K}), we find
\bea
{\cal E} & = &  - \int_C k \cdot \eps + \int_\infty (\bK - t \cd \bB)
\nonumber \\
& = & \int_C T_{ab} n^a t^b \tilde{\eps} + \int_\infty (\bK - t \cd \bB)
\label{Tab}
\eea
where $n^a$ denotes the future-directed unit normal to $C$, and
$\tilde{\eps}_{b_1...b_{n-1}} = n^a \eps_{ab_1...b_{n-1}}$
is the natural volume
element on $C$. Thus, apart from some possible surface term contributions
which can arise from both $\bK$ and $\bB$, the
canonical energy is given by the usual formula involving an
integral of the stress-energy tensor over $C$.

As noted above, in general ${\cal E}$ is not conserved, i.e., independent
of choice of Cauchy surface, $C$. However, if the
spacetime metric is
stationary, $\Liet g_{ab} = 0$, (but stationarity need {\em not} be
imposed upon the matter fields), then the first line of eq.(\ref{cons})
shows that $\bJ$ is closed. Equation (\ref{K}) then immediately implies
that the stress-energy current form
$- k \cdot \eps$ also is closed (as also could easily be verified directly).
Equation (\ref{E5}) then implies that
${\cal E}$ does not change when the Cauchy surface,
$C$, undergoes variations of compact support. In the usual case
where $t \cd \bB$ vanishes at infinity and $\bJ$ goes to zero
suitably rapidly at infinity, ${\cal E}$ will take the same
value for all asymptotically flat Cauchy surfaces.

Now, suppose that $g_{ab}$ is stationary -- i.e., $\Liet g_{ab} = 0$ --
and suppose that $\psi(\lambda)$ is a one-parameter family
of solutions to the matter equations of motion (in the
fixed metric $g_{ab}$)
such that $\psi(0)$ is stationary, i.e., $\Liet \psi(0) = 0$. Let
${\cal E} (\lambda)$ denote the canonical energy of these solutions. Then,
since $\Liet g_{ab} = 0$,
by eqs.(\ref{dH}) and (\ref{B}), we have for all $\lambda$
\be
\frac{d {\cal E}}{d \lambda} (\lambda) =
\Om (\psi (\lambda), \frac{d \psi}{d \lambda}, \Liet \psi(\lambda))
\label{dE}
\ee
In particular, since $\Liet \psi$ vanishes at $\lambda = 0$, we see that
the first variation of the canonical energy about a stationary solution
vanishes,
\be
\de {\cal E} = 0
\ee
Now take the derivative of eq.(\ref{dE}) with respect to $\lambda$ and
evaluate the resulting
equation at $\lambda = 0$. A nonzero contribution will occur on
the right side only when the $\lambda$-
derivative acts on $\Liet \psi$. We thereby find that the second variation
of canonical energy about a stationary solution $\psi$ is given by
\be
\de^2 {\cal E} \equiv \frac{1}{2} \frac{d^2 {\cal E}}{d \lambda^2}
|_{\lambda = 0} = \frac{1}{2} \Om (\psi, \de \psi, \Liet \de \psi)
\label{EOm}
\ee
where
\be
\de \psi \equiv \frac{d \psi}{d \lambda} |_{\lambda = 0}
\ee
Note, in particular, that $\de^2 {\cal E}$ depends only upon $\de \psi$,
and not upon $\de^2 \psi$.

Equation (\ref{EOm}) is one of the key results of this Appendix. To
elucidate its meaning, we
note that if $\de \psi$ satisfies the linearized equations of motion about a
stationary solution, $\psi$, then so does $\Liet \de \psi$. Consequently,
the symplectic current form $\om (\psi, \de \psi, \Liet \de \psi)$,
defined above by eq.(\ref{omega}), is closed. Thus, its integral over a
Cauchy surface, $C$, yields a conserved quantity for perturbations.
Equation (\ref{EOm}) shows that, apart from a factor of $2$,
this conserved quantity is just the second
order change in the canonical energy associated with this perturbation.
By our previous results, we see that
this conserved quantity is equivalent -- up to
possible ``surface terms" -- to the
conserved quantities $\int_C \de^2 \bJ$ and
$\int_C \de^2 T_{ab} n^a t^b \tilde{\eps}$.

As a simple application of the above result, consider the theory of a linear
field $\psi$ in a stationary spacetime, where by ``linear" we mean that
$\bL$ is quadratic in $\psi$, so that the equations of motion for $\psi$ are
linear. In this case, the equations of motion are the same as the
linearized equations about
$\psi = 0$, so we may choose the ``unperturbed solution" to be
$\psi = 0$, and we may write $\delta \psi$ = $\psi$ in the above
formulas. We also have $\de^2 {\cal E} = {\cal E}$
and $\de^2 T_{ab} = T_{ab}$.
Hence, we obtain from eqs.(\ref{Tab}) and (\ref{EOm})
\be
 \Om (\de \psi, \Liet \de \psi) =
2 \int_C T_{ab} n^a t^b \tilde{\eps} + 2 \int_\infty (\bK - t \cd \bB)
\label{EOm'}
\ee
Again, for the types of theories usually considered (such as a Klein-Gordon
scalar field), the surface
terms from infinity in eq.(\ref{EOm'}) vanish. The resulting
relation plays an important role in defining a natural vacuum state
for linear quantum fields in a stationary spacetime \cite{AM}
\cite{Kay}.

Consider, now, the case
where the nondynamical metric is a flat metric, $\eta_{ab}$. We denote
the (flat) derivative operator associated with $\eta_{ab}$ by
$\partial_a$. Let $\k^a$ be a translational Killing field of $\eta_{ab}$, so
that $\partial_a \k^b = 0$. Then, clearly, at each point of spacetime
the Noether current $\bJ$ associated with
$\k^a$ is linear in the value of $\k^a$ at that point.
Hence, there exists a unique tensor field, ${{\cal T}^a}_b$, called the
{\em canonical energy-momentum tensor}, such that
\be
J_{a_1...a_{n-1}} = - {{\cal T}^a}_b \k^b \epsilon_{a a_1...a_{n-1}}
\label{calT}
\ee
Conservation of $\bJ$ implies conservation of ${{\cal T}^a}_b$ in its
first index, i.e.,
\be
\partial_a {{\cal T}^a}_b = 0
\ee
However, ${\cal T}^{ab}$ need not be symmetric.
Nevertheless, eq.(\ref{K})
implies that there exists a tensor field $H^{abc} = H^{[ab]c}$,
locally constructed out of $\eta_{ab}$ and $\psi$, such that
\be
{\cal T}^{ab} = T^{ab} + \partial_c H^{cab}
\label{Habc}
\ee
Thus, we have re-derived the well known fact
that ${\cal T}_{ab}$ always can
be ``symmetrized" by the addition of an
identically conserved tensor $\partial_c H^{cab}$.

Finally, we note that much of the theory of pseudotensors can
be derived by applying the results of this Appendix
back to the case where the metric
again is a dynamical variable in the Lagrangian (\ref{new1'}). For a
diffeomorphism invariant theory of the type considered in the body
of this paper, we may
introduce a fixed flat metric, $\eta_{ab}$, on
spacetime, and express the dynamical metric, $g_{ab}$, as,
\be
g_{ab} = \eta_{ab} + h_{ab}
\label{g}
\ee
We then may treat $\eta_{ab}$ and $h_{ab}$ as independent fields,
and view our theory as a theory with Lagrangian of the form
(\ref{new1'}) with the dynamical fields
$(h_{ab}, \psi)$ and a nondynamical metric $\eta_{ab}$. One of the
(very few) advantages
of doing this is that many more quantities qualify as ``covariant" when
$\eta_{ab}$ and $h_{ab}$ are viewed as independent fields. In particular,
in general relativity, no diffeomorphism covariant $(n-1)$-form, $\bB$,
satisfying eq.(\ref{defB}) can be constructed out of $g_{ab}$, but there
is no difficulty in constructing a diffeomorphism covariant $\bB$ out of
the independent fields $\eta_{ab}$ and $h_{ab}$.
Thus, we may change the Lagrangian via
\be
\bL \rightarrow \bL' = \bL - d \bB
\label{varL}
\ee
and still view $\bL'$ as being of the general form (\ref{new1'})
(with $\eta_{ab}$ and $h_{ab}$ viewed as independent fields).
Under the change of Lagrangian (\ref{varL}), $\th$
is modified by
\be
\th \rightarrow \th' = \th  - \delta \bB
\ee
(see eq.(\ref{chth}) above). Consequently, we have
$\int_\infty t \cd \th' = 0$, and the canonical energy
now is given by simply
\be
{\cal E} = \int_C \bJ'
\label{J'}
\ee

Since the theory has been recast to have a Lagrangian of the form
(\ref{new1'}) in a spacetime with a nondynamical flat metric
$\eta_{ab}$, a canonical energy-momentum tensor can be defined
by eq.(\ref{calT}). We denote this tensor as ${t^a}_b$,
and refer to it as a
{\em pseudotensor} because it
depends upon the choice of flat metric, $\eta_{ab}$
and thus is not covariant with respect to diffeomorphisms which act
only upon the dynamical fields. For the case of
vacuum general relativity with the Lagrangian $\bL'$ of eq.(\ref{varL})
with an appropriate choice of $\bB$,
${t^a}_b$ corresponds to the Einstein pseudotensor \cite{So}.
Note that eq.(\ref{J'}) can be rewritten in terms of ${t^a}_b$ as,
\bea
{\cal E} & = & \int_C {t^a}_b n_a t^b \tilde{\eps} \nonumber \\
& = & \int {t^0}_0 d^3 x
\label{canen}
\eea
where the last line holds when $C$ is taken to be the hypersurface
$t = constant$ in a global inertial coordinate system of $\eta_{ab}$.

In order to define a stress-energy tensor corresponding to (\ref{defT}),
we must specify the functional dependence of
the Lagrangian on $\eta_{ab}$
for general (non-flat) $\eta_{ab}$. One way to do this would be to take
$\bL$ for general $\eta_{ab}$ to be given by the substitution (\ref{g})
in the original Lagrangian. In that case, $\bL$ clearly depends upon
$\eta_{ab}$ and $h_{ab}$
only in the combination $\eta_{ab} + h_{ab}$. Consequently,
the equations of motion for $\eta_{ab}$ will be satisfied whenever
the equations of motion for the dynamical field $h_{ab}$ hold. Note that
$\bB$ need not depend only on the combination $\eta_{ab} + h_{ab}$,
so $\bL'$, defined by eq.(\ref{varL}), need not depend only upon this
combination. Nevertheless, since addition of an exact form to $\bL$
does not alter the equations of motion, it remains true for $\bL'$
that the equations of motion for $\eta_{ab}$ will be satisfied whenever
the equations of motion for the dynamical field $h_{ab}$ hold.
But, this implies that the
energy-momentum tensor defined by (\ref{defT})
vanishes by virtue of the equations of motion for
the dynamical fields. Equation (\ref{Habc}) then yields,
\be
t^{ab} =  \partial_c H^{cab}
\label{H8}
\ee
This proves that -- when the equations of motion are imposed --
the pseudotensor ${t^a}_b$ always can be derived
from a ``superpotential" $H^{cab}$. Consequently, the volume integral
(\ref{canen}) always can be converted to a surface integral at infinity.
This fact, of course, corresponds to our previous result, eq.(\ref{E}),
which was derived in a much more simple and direct manner.

When recast in the form (\ref{new1'}), the Lagrangian
obtained from $\bL'$ by
the simple substitution (\ref{g}) described above will, in general,
have a nontrivial, explicit dependence upon the
curvature of $\eta_{ab}$. However, an alternative procedure
for defining a Lagrangian for non-flat $\eta_{ab}$ -- which clearly
agrees with $\bL'$ when $\eta_{ab}$ is flat -- would be to modify
the Lagrangian of the previous paragraph by simply setting the
terms in $\bL'$ involving the curvature of $\eta_{ab}$ to zero. If we
do so, the stress-energy tensor
defined by (\ref{defT}) for this
modified Lagrangian will be nonvanishing. We denote this stress-energy
tensor by $\tilde{t}_{ab}$. If -- as seems plausible --
the surface term $\bK$ does not contribute to eq.(\ref{Tab}), then
the symmetric pseudotensor $\tilde{t}_{ab}$ will be equivalent to
${t}_{ab}$ insofar as the calculation of canonical energy is concerned,
i.e., eq.(\ref{canen}) will hold with ${t}_{ab}$ replaced by $\tilde{t}_{ab}$.
Note that eqs.(\ref{Tab}) and (\ref{H8}) imply that $\tilde{t}_{ab}$ also is
derivable from a superpotential.
Other symmetric pseudotensors derivable from a superpotential
can be explicitly constructed in the case of general relativity
(see, e.g., \cite{LL}).

The dependence of pseudotensors such as ${t}_{ab}$ or $\tilde{t}_{ab}$
on the choice of $\eta_{ab}$ significantly
limits their physical interpretation and utility. Indeed, it is difficult
to imagine any use to which they could be put other than for the definition or
calculation of the canonical energy and other asymptotic conserved quantities
-- and this can be accomplished much more straightforwardly
by the methods described in the body of this paper.
Nevertheless, the canonical energy can be correctly computed from
a pseudotensor via eq.(\ref{canen}).
In particular, if we consider perturbations of a stationary solution and
choose the flat
metric $\eta_{ab}$ so that the Killing field, $t^a$, of the stationary
background is a translational Killing
field of $\eta_{ab}$, then
\be
\delta^2 {\cal E} = \int \delta^2 {t^0}_0 d^3 x
\ee
is a nontrivial conserved quantity which depends only on the first
order perturbation of the dynamical fields.
For an arbitrary pseudotensor, this formula for $\delta^2 {\cal E}$ is not
very useful because to get an expression for the conserved quantity
in terms of the first order perturbation, one must use the second order
field equations to eliminate the terms in
$\delta^2 {t^0}_0$ which involve the
second order perturbation. However, as shown by Sorkin \cite{So}, for the
Einstein pseudotensor, the terms in $\delta^2 {t^0}_0$ involving the second
order perturbation are separately conserved (irrespective of the second
order field equations), so the Einstein pseudotensor can be used to obtain
a nontrivial conserved quantity constructed out of the first
order perturbation of the dynamical fields.
For perturbations of static,
electrovac spacetimes in general relativity, this
conserved quantity is equivalent
to the conserved flux integral obtained by Chandrasekhar and Ferrari
\cite{CF1,CF2}. As was shown explicitly in
\cite{BW}, the Chandrasekhar-Ferrari conserved current also is
equivalent to the symplectic current $(n-1)$-form
$\om (\phi; \de \phi, \Liet \de \phi)$, which directly yields
$\delta^2 {\cal E}$ by eq.(\ref{EOm}) above. Note that $\om$ is
constructed entirely out of the dynamical
fields and their perturbations -- in particular, no background flat metric
need be introduced -- so it provides a covariant version of the
conserved flux integral. However, $\om$ is not gauge invariant under
infinitesimal gauge transformations of the perturbed dynamical fields, so
it also does not provide a meaningful notion of the
local energy density of the perturbation.

\end{section}

\end{document}